\PassOptionsToPackage{hyphens}{url}

\documentclass{packages/icmc}

\usepackage{rotating}           
\usepackage[all,knot,arc,import,poly]{xy}   

\usepackage{times,newtxmath}       
\usepackage[cal=cm,scr=boondoxo,bb=ams]{mathalfa} 
\usepackage{float} 

\DeclareMathOperator*{\argmax}{arg\,max}

\usepackage{amsthm}
\newtheoremstyle{exampstyle}
  {} 
  {} 
  {} 
  {} 
  {\bfseries} 
  {} 
  {.5em} 
  {{\thmname{#1}\thmnumber{ #2}\textbf{\thmnote{ (#3)}}}} 

\usepackage{array} 

\theoremstyle{exampstyle}

\usepackage{inconsolata}

\usepackage{natbib}         

\usepackage{etoolbox}

\allowdisplaybreaks

\newcommand{\overbar}[1]{\mkern 1.5mu\overline{\mkern-1.5mu#1\mkern-1.5mu}\mkern 1.5mu}

\tituloPT{Modelos Probabilísticos para o Tempo de Execução de Tarefas Individuais em Escalonamento Estocástico}
\tituloEN{Probabilistic Models for the Execution Time in Stochastic Scheduling}
\autor[Saldanha, M. H. J.]{Matheus Henrique Junqueira Saldanha}
\genero{M} 
\orientador{Prof. Dr.}{Adriano Kamimura Suzuki}
\curso{CCMC}
\data{07}{06}{2020} 
\idioma{EN} 

\textoresumo[english]{
    The execution time of programs is a key element in many areas of computer science, mainly those where achieving good performance (e.g., scheduling in cloud computing) or a predictable one (e.g., meeting deadlines in embedded systems) is the objective. Despite being random variables, execution times are most often treated as deterministic in the literature, with few works taking advantage of their randomness; even in those, the underlying distributions are assumed as being normal or uniform for no particular reason. In this work we investigate these distributions in various machines and algorithms. A mathematical problem arises when dealing with samples whose populational minimum is unknown, so a significant portion of this monograph is dedicated to such problem. We propose several different effective or computationally cheap ways to overcome the problem, which also apply to execution times. These methods are tested experimentally, and results point to the superiority of our proposed inference methods.
    We demonstrate the existence of execution time distributions with long tails, and also conclude that two particular probability distributions were the most suitable for modelling all execution times. While we do not discuss direct applications to stochastic scheduling, we hope to promote the usage of probabilistic execution times to yield better results in, for example, task scheduling.
}{execution time, stochastic scheduling, probability distribution, statistical modelling, worst case execution time}

\incluilistadefiguras

\incluilistadetabelas

\incluilistadesiglas

\begin{document}
\urlstyle{tt}

\renewcommand\rev[1]{}
\renewenvironment{revenv}{\comment}{\endcomment}
\allowdisplaybreaks

\textual

\chapter{Introduction}
\label{chapter:intro}

The execution time of programs is a key element in scheduling problems, which often appear in the context of cloud computing, distributed and parallel programming, as well as in operating systems (especially real-time systems) \citep{tanenbaum2015modern,bittencourt2012scheduling}. However, execution time is a random variable, in the sense that running the same program multiple times yields different execution times; yet, as is discussed in this monograph, it appears that the literature in scheduling for the cloud heavily neglects this randomness, by facing execution times as deterministic or considering only the average time to perform the scheduling. Furthermore, the complexity of computer architectures has been growing continuously, incorporating deeper pipelines, larger cache hierarchies, more involved speculation and prediction techniques, support for more levels of parallelism, and so forth\footnote{Techniques used by modern architectures are thoroughly explained by \citet{hennessy2011computer}.}. Many of such techniques contribute to increase the unpredictability of computer programs' execution times, which poses a problem to scheduling.

Consider the \textit{scheduling problem}, for example, which takes a \sigla{DAG}{directed acyclic graph} of tasks $T_1, \dots, T_n$ and tries to determine the optimal way to execute them in the available processors in order to finish all the tasks in the lowest possible time. The objective is thus to minimize the \textit{makespan}, the overall execution time for such DAG\footnote{Other objectives can be formulated, but here we limit ourselves to this one.}; to do this, we must answer the question \textbf{what is the optimal order to execute these tasks in the available resources?} This problem appears naturally in cloud computing (e.g., when planning the execution of a workflow provided by the user \citep{yu2007multi, bittencourt2012scheduling}) and parallel computing (e.g., load balancing in parallel algorithms \citep{foster1995designing, li1997estimating}). The scheduling problem has often been approached in a deterministic way, using the average execution time of tasks to make decisions. Some probabilistic models have also been proposed for the problem (in this case it is called \textit{stochastic scheduling}\footnotemark), mostly under the assumption that the probability distribution of tasks' times is known, such as done by \citet{li1997estimating} and \citet{zheng2013stochastic}. In both of these works, in order to validate their results, the authors performed simulations where the the distributions of tasks are somewhat arbitrarily chosen to be either normal or uniform; as will be discussed in Section \ref{sec:cloud-computing}, this seems to be common practice in the literature despite there being no scientific evidence that the underlying distributions are indeed normal or uniform. In the present work, we further investigate whether these are reasonable choices of distributions for execution times, and conclude that in some cases they are not.
\footnotetext{This is not to be confused with scheduling algorithms that make use of stochastic optimization approaches such as particle swarm or genetic algorithms. These can be used to minimize the makespan, regardless of whether the makespan involves deterministic execution times or random ones.}

The applicability of the results presented in this monograph are most evident in the problem of stochastic scheduling for minimizing the makespan. However, it is worth commenting on two related problems that can also benefit from our results. One is the area of real-time systems, in which programs are studied whose response time is critical. In the specific case of \textit{hard} real-time systems \citep{tanenbaum2015modern} it is crucial that deadlines be met, and failing to do so may result in casualties or other catastrophes (e.g., embedded systems within modern cars). To ensure that deadlines are met, one has to determine the worst-case execution time (WCET, see \citet{wilhelm2008worst}) of the software in question, which has received a decent amount of attention from the scientific community and is discussed in Section \ref{sec:wcet}.

The second related area is that of high performance computing in general, which includes parallel and distributed computing \citep{rauber2013parallel}, compiler optimization \citep{tripathy2015dynamic} and algorithms in general \citep{cormen2009introduction}. As it is concerned with accelerating existing programs, papers published in this area often involve comparing the execution times of two programs, the existing and the proposed ones (e.g., \citet{soman2010fast}, \citet{zhang2013parallel}). Naturally, a hypothesis test can be applied to validate such comparison, which in turn relies in assumptions about the underlying probability distribution of the execution times. If it is sufficiently similar to a normal distribution, a sample of size around 30 is enough to justify approximating the sample mean by a normal distribution \citep{walpole1993probability}. However, rather than taking this as a well-established golden number, it might be of interest to investigate what sample sizes are satisfactory, which requires knowledge of the underlying probability distribution.

In this ongoing project we investigate the underlying distribution of execution times of programs, attempting to reason about: 1) the suitability of varied probabilistic models for execution times; 2) the causes for the observed randomness; 3) which shapes of probability distributions are observed and how this affects scheduling; and 4) methods to deal with problems faced while investigating the previous points. The main hindrance we found was performing maximum likelihood estimation over variables whose ``location'' (or populational minimum) is large and unknown, which we have thoroughly investigated and thus far proposed six inference methods that are discussed and justified in Chapter \ref{chapter:inf-location-parameter}. On top of that, experiments were performed to test the capacity of a specified set of distribution families to fit samples of execution times obtained experimentally, which is discussed in Section \ref{sec:experiments-results}. Results demonstrate the existence of execution time distributions with heavy tails on either side, which could indeed have significant impact in scheduling decisions. We show that the exponentiated Weibull and the \sigla{OLL-GG}{odd log-logistic generalized gamma} distributions achieve very good performance when fitting samples of execution times. Chapter \ref{chapter:context} provides the context in which this work lies, elaborating on areas where execution times assume a central role, commenting on existing work on such areas, and also presenting related work that have already attempted to take advantage of treating execution times as random variables. Chapter \ref{chapter:methods} is dedicated to explain concepts that are used throughout the text but might be unfamiliar to the reader. The aforementioned results and details specific to execution times (and not the general setting) are discussed in Chapter \ref{chapter:unfolding}. Finally, the monograph is concluded in Chapter \ref{chapter:conclusion}.

\chapter{Context and Related Work}
\label{chapter:context}

In this chapter we elaborate on the areas where execution times assume a key role, namely cloud computing, distributed and parallel computing and embedded systems. We mentioned previously that our results have an impact on how hypothesis tests are made in computer science, but this subject is explored only in the next chapter as it weights more on the theoretical side. Along the development of this chapter, a myriad of related work from the literature is also referenced and commented.

\section{Cloud Computing}
\label{sec:cloud-computing}

Cloud computing is the use of outsourced computational resources on-demand in a pay-per-use basis, and has been made feasible thanks to advances in operating systems and techniques for virtualization \citep{tanenbaum2015modern}. In this business model, a single entity holds a large amount of computer machines (the \textit{cloud}), and offers these machines for use by interested users. It has become financially interesting because extreme computational power is needed only occasionally in many cases. With the cloud, instead of buying powerful machines that might become idle most of the time, businesses can leverage the necessary computational power from the cloud solely for the time needed.

In general, users are able to send jobs to execute in the cloud, which are often represented as a directed acyclic graph (DAG) (in this context, the graph is commonly called a \textit{workflow}) where each node represents a computational task that produces a certain data, and edges represent data dependencies between these tasks. A task cannot be started before all data dependencies have been met. Finally, users can submit multiple workflows, and tasks in workflows of different users can be mapped to the same machine.

There are three main service models offered in the context of cloud computing, which assume the so called acronyms \sigla{IaaS}{infrastructure as a service}, \sigla{PaaS}{platform as a service} and \sigla{SaaS}{software as a service} \citep{coulouris2005distributed}. In IaaS the client has full control of the computational environment, and has to specify the desired amount of storage, CPU power, memory, network bandwidth and other hardware-specific aspects; the client also has to perform due configurations in the operating system to correctly use the requested hardware. PaaS clouds are often oriented to specific applications (e.g., website server), so they offer cloud plans with fixed hardware specifications, which are tailored to suit the application approximately well; these platforms also come with a pre-configured operating system that is ready to run the application. Finally, SaaS offers a software that can be used by the user through a web interface, and the software runs on an underlying machine that is not available for the user to modify or configure; Google's GSuite\footnote{See \url{https://gsuite.google.com/} (access in September 26th, 2019).} is an example of this model. In the literature, there does not seem to exist an agreement concerning precise definitions of IaaS, PaaS and SaaS, but the general ideas are as discussed above.

One difficult problem in cloud computing, which has received a lot of attention from the scientific community, is the problem of scheduling workflows in the available machines \citep{bittencourt2012scheduling}. As mentioned previously, users can submit multiple workflows, which in turn are composed of a graph of interconnected tasks; the main question that arises here is \textbf{what is the optimal order to execute these tasks in the available resources?} This is the problem of scheduling, which is known to be NP-complete in many of its forms \citep{ullman1975np}, most of which are formulated as a problem of multi-objective optimization (as done, for example, by \citet{rodriguez2014deadline} and \citet{zuo2015multi}). There are many objectives that can be pursued here, such as minimizing the makespan (overall execution time of a workflow), minimizing the cost for the client, maximizing the throughput of completed workflows, and so on. Due to the difficulty in solving the problem analytically, approximate or heuristic optimization methods are frequently used \citep{zhan2015cloud, bittencourt2012scheduling}.

In the literature, most proposed scheduling algorithms consider only the mean execution time of the tasks in order to decide the mapping, rather than other information such as the variance or the probability distribution itself. For example, \citet{panda2015efficient} propose two algorithms that require a matrix of expected execution times of each task $T_i$ on each machine $M_j$. \citet{rodriguez2014deadline} applies \sigla{PSO}{particle swarm optimization} for scheduling, and it also uses the mean execution time of tasks, though automatically calculated based on: 1) the capacity of each machine (in FLOPS) and 2) the size of the task (in number of operations). Similarly, \citet{zuo2015multi} apply ant colony optimization and, besides the mean time, no other information concerning time is used. The same happens in the works of \citet{tsai2014hyper}, \citet{tang2015self}, \citet{tripathy2015dynamic} and \citet{xavier2019chaotic} (although the latter also uses the standard deviation of the processing load on each machine, which is quite interesting).

This phenomenon is more than expected, considering the difficulty of the problem: authors place their focus on complications such as the heterogeneity of machines in the cloud, the necessity for complying with deadline and \sigla{QoS}{quality of service} requirements, the use of task preemption to improve results, etc. Simplifying the model -- by assuming that tasks' execution time can be characterized by a single number (the mean) -- allows simpler algorithms to be devised, yet achieving outstanding results. It is arguable, however, that most of these proposed algorithms can be modified to make use of the randomness of execution times, and consequently improve results, as is discussed in this monograph. We demonstrate that execution times demonstrate a relatively complex behavior across different machines and program types, so it seems that the mean is not sufficient to decide the mapping of tasks into machines.

The area of stochastic scheduling is responsible for using more information about execution times of particular tasks\footnote{Recall that workflow is a graph of tasks.} (often statistical information) to perform the scheduling, and there has been a decent amount of work in such subject, although much less than in the deterministic context. \citet{li1997estimating} provide a general statistical framework way to calculate the probability distribution of the makespan of a generic graph of tasks, be in the context of cloud computing or not; it served as a basis for subsequent stochastic scheduling algorithms. This is not a trivial problem because it can involve a maximum of two non-independent random variables, which is mathematically cumbersome; the authors smartly use the Kleinrock independence approximation from queuing theory \citep{kleinrock2007communication} to mitigate this problem and allow calculations. Their approach requires the time distribution of tasks to be known, which for experimentation purposes they assume as being normal distributions; one interesting implication of our work is to provide more informed suggestions as to which distributions could be used. An approach for predicting the execution time of a task given its previous behavior ``features'', is given in \citet{ganapathi2010statistics}. The authors perform experiments specifically with map-reduce jobs \citep{chu2007map}, but the idea of correlating previous behavior, such as average time for transferring data or the amount of data processed, is promising and can be extended to contexts within cloud computing.

Heuristic optimization methods whose objective function involve execution times modelled as random variables are presented in \citet{dogan2001stochastic} and \citet{dogan2004genetic}. \citet{shestak2008stochastic} proposes a stochastic metric for evaluating the robustness of a scheduling algorithm, which is given by the probability that the time will be in some defined interval $[\beta_1, \beta_2]$, which requires the probability distribution of tasks to be given. Here the authors claim that normal distributions happen frequently in practice; on that note, \citep{chen2016uncertainty} proposes a stochastic scheduling algorithm that assumes all execution times to be normal. \rev{Should this be here? ->} \citet{kamthe2011stochastic} use, besides the mean, the standard deviation of execution times in order to improve scheduling results. A method for estimating an empirical probability density function for execution times is given in \citet{dong2010resource}, which involves estimating histograms and, therefore, require a decent number of available samples. \citet{cai2017elasticsim} presents a framework for simulating scheduling algorithms which support stochastic execution times, and the requirement is that the probability distribution be defined by the user. Finally \citet{zheng2013stochastic} uses Monte Carlo simulation to overcome the mathematical difficulty of stochastic scheduling, but it also expects underlying probability distribution of tasks to be given, and for experimentation purposes they assumed normal or uniform distributions.

\rev{We could use profiling results to justify the normality of the probability distributions. If the program is big, but it passes 98\% of the time in a single procedure, then it probably is not normal. If the time is more evenly spread among procedures, then it is more likely to be normal.}

\section{Extension to Distributed and Parallel Programs}
\label{sec:extension-to-distributed-programs}

The problem of scheduling is also present in the areas of distributed and parallel computing, with small modifications from the problem presented in the previous section. Let us first state the difference between these two areas: we will refer as distributed any program that explores parallelism by leveraging multiple computers (each computer having a single operating system) that do not share memory (e.g., cluster of computers or a supercomputer \citep{tanenbaum2016structured}). Parallelism in single multicore machine, with shared memory or not, and possibly with coprocessors such as GPUs, will be considered in this text as parallel computing.

Considering these definitions, we see that a cloud is a distributed system; consequently, their scheduling algorithms have a large intersection. However, cloud computing involves a specific flow of job submission and job execution, so scheduling algorithms often take this in consideration to map tasks into machines. In contrast, there are other distributed systems with different characteristics; for example, a shared web hosting machine that runs web servers of many users, each of which may interact with databases located in other (distant) machines, would require other scheduling strategies.

In the case of parallel algorithms, the time for moving data between machines tends to be shorter and get comparable to the time for processing such data. One way to schedule tasks in a parallel program is to model these times with computational models \citep{norman1993models, rauber2013parallel}, in which the time for moving data or executing a task is considered to be a deterministic function of parameters such as the size of data to be transferred or the dimensions of the input data.

As an example, take the problem of training a neural network. This involves computing gradients such as the following
$$
\sum_{(x, y) \in \mathcal{D}} \nabla_{W_f} L(y, h(g(f(x, W_f), W_g), W_h))
$$
where $\mathcal{D}$ is our dataset containing pairs $(x, y)$ (the data item and its correct label, respectively), $W_i$ are weights associated with some function $i$, $L$ is the loss function that will tell how ``wrong'' the neural network currently is, and $f, g, h$ are layers (e.g., fully-connected, convolutional, max-pooling) that apply some non-linear transformation to their inputs \citep{goodfellow2016deep, haykin1994neural}. The sum can clearly be split over disjoint subsets of $\mathcal{D}$, and each of these splits can be calculated in a different processor, core or coprocessor. Considering a cluster with 12 computers, each containing a multicore processor with 2 cores and one GPU, we would like to have $12 \cdot (2 + 1) = 36$ splits to compute in each core and GPU. Further, we would like that each split be computed in the same time, so the splits must have different sizes (as GPUs will likely be much faster \citep{rauber2013parallel}); how can we calculate the size of each split? The results presented in this monograph could be used to help such decision\footnote{This whole problem is part of an ongoing project; see \url{https://mjsaldanha.com/sci-projects/2-accel-nns-1/}.}.

\section{Worst-Case Execution Time in Embedded Systems}
\label{sec:wcet}

Time is a paramount factor in many embedded systems, namely real-time systems and more specifically the hard real-time systems \citep{tanenbaum2015modern}. Here, deadlines must be followed by the underlying program, otherwise catastrophes could happen; take, for example, the \sigla{ABS}{anti-lock braking system} that controls the brakes of a car. Since the execution time is a random variable, and considering that deadlines must \textit{always} be followed, it is necessary to calculate the \sigla{WCET}{worst-case execution time}, that is, the largest time that the program can take when in the production environment.

One way to go about solving this problem is to estimate the underlying probability distribution, and the worst-case scenario will be on the right-tail thereof \citep{wilhelm2008worst}. With the distribution in hand, the quantity of interest is $t$ for which $P(T > t) \approx 0$, which is the time above which events are extremely rare or impossible. However, estimating the underlying distribution is far from trivial, and the first difficulty comes from the fact that input data (e.g., data from sensors interacting with the environment) is not deterministic, and each tuple of inputs may cause the program to follow very different execution paths in its \sigla{CFG}{control flow graph}, which in turn may greatly change the generation process of execution times.

As will be more elaborated in Section \ref{sec:problem-program-inputs}, it is reasonable to say that a single execution path in the CFG yields execution times following a certain well-behaved\footnote{Consider as well-behaved a smooth distribution with low variance and few modes.} distribution, but the same is not true between different paths. In other words, the distribution given an input can be expected to be well-behaved, but the distribution averaged over all possible inputs may be very uninformative. The program would then have to be described by multiple probability distributions, one for each set of similar execution paths, then each distribution could be analyzed separately (e.g., \citet{david2004static}). Alternatively, one could define a probability distribution for the inputs given to the program, which allows merging the multiple probability distributions of paths into a single distribution, giving more weight to the paths that are more often exercised (e.g., \citet{braams2016deriving}).

There is a lot of existing work in the area of WCET: a survey is given in \citet{wilhelm2008worst}. One common approach is to find WCET statically, by taking the source or compiled binary code, building its control flow graph and analyzing its behavior on a certain model of processor. In \citet{david2004static}, a framework for calculating WCET statically is presented, which provides the user with the WCET for a given execution path in the CFG, as well as the probability for the program to follow such a path; however, it requires the user to provide probability distributions for many aspects of the program and its environment, which is a shortcoming of the method. Similar works can be found in \citet{ferdinand2004ait}, \citet{wilhelm2010static}, \citet{engblom2002processor}, \citet{li1995performance} and \citet{li2007chronos}. Caches are known to be a hard part of static analysis, as it has the biggest impact in execution times \citep{hennessy2011computer}, so it has received special attention in \citet{lv2016survey}, \citet{patil2004compositional}, \citet{touzeau2017ascertaining}, \citet{blass2017write} and \citet{kosmidis2013multi}. In particular, \citet{kosmidis2013multi} discusses one way to deal with the problem of statically predicting the number of cache hits or misses: the authors propose and discuss a cache design that is randomized, in the sense that the eviction and replacement of cache block is not done in usual ways (e.g., the \textit{least recently used} and \textit{write back} approaches), and then they provide proofs that such a cache greatly favors WCET analysis.

Another way to approach WCET is by using execution time measures taken by running the software on the given hardware using a specified set of inputs \citep{wilhelm2010static}. This problem is often formulated probabilistically: let $X_1, \dots, X_n$ be an \sigla{iid}{independent and identically distributed} random sample of execution times, then the problem is the estimation of $M_n = \max\{ X_1, \dots, X_n\}$. The probability distribution of $M_n$ can be calculated analytically:
\begin{align*}
    P(M_n < t) &= P(X_1 < t, \dots, X_n < t) \\
    &= P(X_1 < t) \cdots \dots \cdot P(X_n < t) \\
    &= F(t)^n
\end{align*}
where $F(t)$ is the cumulative distribution of the variables $X_i$. This requires, however, knowledge of $F(t)$ or, equivalently, the underlying probability distribution of the variables $X_i$, which is often not available. Thankfully, the area of \sigla{EVT}{extreme value theory} \citep{coles2001introduction} has important results regarding the distribution of the sample maximum even if $F$ is unknown. The Fisher–Tippett–Gnedenko theorem \citep{fisher1928limiting, gnedenko1943distribution} says that, in the limit when the sample size goes to infinity, the distribution of $(M_n - b) / a$ with $a, b \in \mathbb{R}$ converges (when it converges to a non-degenerate distribution) to one out of three distribution families: Gumbel, Fréchet or Weibull.

Many works explore this result in measure-beased WCET, such as done by \citet{lima2016extreme}, \citet{abella2017measurement}, \citet{silva2017using} and \citet{lu2011new}. However, it often happens in WCET that the assumption of the theorem does not hold, as empirically observed in \citet{lima2016extreme}, though in the same work the authors comment on some cases where EVT does indeed help determining the WCET. Besides EVT there are other ways to approach the measure-based WCET problem. \citet{lindgren2000using}, for example, uses code instrumentation to trace the program's execution path indirectly, which leads to a system of equations whose solution helps determining the WCET. Measure-based approaches are also used to improve estimates of the static approaches, as done in \citet{wenzel2005measurement} and \citet{kirner2005using}.

\rev{Maybe check \url{https://www.tandfonline.com/doi/full/10.1080/08874417.2016.1180651}, \url{https://www.sciencedirect.com/science/article/pii/S0268401216301980}, \url{https://www.sciencedirect.com/science/article/pii/S0747563216302412}.}




\chapter{Methods, Techniques and Theoretical Foundation}
\label{chapter:methods}

Besides concepts presented in Chapter \ref{chapter:context}, one concept that is heavily used in this project is that of maximum likelihood estimation (MLE), which is explained in the following. In Section \ref{sec:hypothesis-tests} we elaborate on hypothesis testing which, despite being part of the context of the present project, is placed here because it turned out to be a very theoretical and conceptual section.

\section{Estimation and the Maximum Likelihood Estimator}

Inferring characteristics of a population based on a sample is a widely studied problem in the field of statistics. In the formal setting, we have a random variable $X$ with unknown probability density function $f(x)$\footnote{We limit the explanation to the continuous case, as only this case is used throughout the text.}, but we only have access to a finite sample $X_1, \dots, X_n$ taken from this underlying distribution. The sample is a sequence of random variables because every time we take $n$ elements from the population $X$ it will yield a different result $x_1, \dots, x_n$\footnote{We will use the convention that uppercase letters mean random variables, whereas lowercase mean realizations (instantiations) of such random variables.}.

In order to infer characteristics about the population $X$, we can observe the characteristics of the sample. Any function of the sample $X_1, \dots, X_n$ is a potential characteristic of the population, and is thus called an \textit{estimator} \citep{degroot2012probability}. For example, the sample mean defined as
\begin{align*}
    \overbar{X} = \frac{\sum_{i = 0}^n X_i}{N}
\end{align*}
is an estimator for the populational mean. It is a function of random variables, so it is itself a random variable; as such, important quantities such as its expected value and variance can be computed. Given a sample $x_1, \dots, x_n$, the sample mean can be computed and it will deviate from the populational mean; the variance of $\overbar{X}$ tells how much deviation should be expected.

In general, if we have two estimators for the same populational quantity $\theta$, the one with lower variance is called more \textit{efficient}. The estimator $\hat{h}$ is called \textit{consistent} if it converges to the populational quantity as the sample size goes to infinity; rigorously, it is consistent if, and only if
\begin{align*}
    \forall \epsilon > 0 \;\;\; \lim_{n \rightarrow \infty} P( |\hat{h}(X_1, \dots, X_n) - \theta| < \epsilon) = 1
\end{align*}
One last important property of estimators is the bias, defined as
\begin{align*}
    E[\hat{h}(X_1, \dots, X_n)] - \theta
\end{align*}
and the estimator is called \textit{unbiased} if this difference is zero, that is, the estimator's expected value is the same as the estimated populational quantity.

Consider a random variable $X$ whose distribution is not known. However, we know that its distribution comes from a certain family (e.g., normal distribution with parameters $\mu$ and $\sigma$). Therefore, what is left is only to find the parameters of such family of distributions; \textit{maximum likelihood estimator} (MLE) is an estimator for the parameters. Suppose $X$ comes from a distribution with density function $f(x \mid \rho)$ where $\rho$ is the vector of parameters of the distribution family. For an independent and identically distributed (iid) sample $x_1, \dots, x_n$, the probability of having sampled each value is $f(x_i \mid \rho)$, and the probability of having sampled all of them is the product of all these terms, that is,
\begin{align*}
    L(\rho) = f(x_1 \mid \rho) \cdot f(x_2 \mid \rho) \cdot \ldots \cdot f(x_n \mid \rho) = \prod_{i = 1}^n f(x_i \mid \rho)
\end{align*}
where $L$ is called the \textit{likelihood} function. The maximum likelihood estimator for the vector $\rho$ is defined as
\begin{align*}
    \rho = \argmax_{\rho} L(\rho) = \argmax_{\rho} \prod_{i = 1}^n f(x_i \mid \rho) 
\end{align*}
Since the logarithm is a concave function, the same $\rho$ will be found if we apply logarithm on the internal product, that is,
\begin{align*}
    \rho &= \argmax_{\rho} \log \prod_{i = 1}^n f(x_i \mid \rho) \\
    &= \argmax_{\rho} \sum_{i = 1}^n \log f(x_i \mid \rho)
\end{align*}
which is the most common way of performing MLE, as a sum is more numerically stable than a product, avoiding overflows. The MLE is broadly used because it is consistent (in most problems), and its distribution is known under certain conditions to be a normal $N(0, I^{-1} / n)$ where $I$ is the Fisher information (whose definition is out of the scope of this monograph) for $\rho$. This applies if $\rho$ is single parameter; if it is a vector, small modifications are needed \citep{degroot2012probability}.


\rev{Last section discussed about the MLE, where numerical optimization is heavily used, since there is usually no closed form solution for the maximum of the desired expressions.}

\section{Hypothesis Tests}
\label{sec:hypothesis-tests}

Research in computer science, especially in areas related to performance of programs, often involves comparing two populations of execution times. For example, if a novel, faster algorithm $\mathcal{B}$ is being proposed to solve a problem, it is natural to compare its execution time with the usual approach $\mathcal{A}$. This is often done by executing $\mathcal{A}$ and $\mathcal{B}$ multiple times, yielding two sets of samples of execution times, and the mean of the samples are directly used to make statements such as ``algorithm $\mathcal{B}$ is faster than $\mathcal{A}$''. 

\begin{figure}[htb]
    \centering
    \includegraphics[width=0.5\linewidth]{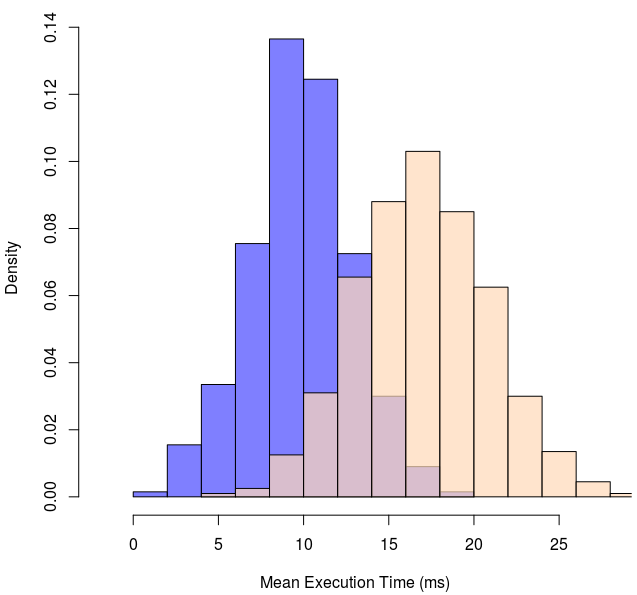}
    \caption{Possible probability distributions for the sample mean of two samples of execution times.}
    \label{fig:hyptest}
\end{figure}

Nevertheless, being random variables, comparing samples of execution times is far from trivial. To start with, let $T$ be the execution time of a program, and $T_1, \dots, T_n$ a sample of such variable (i.e., multiple measures for the same program), then the sample mean defined as
$$
    \overline{T} = \frac{1}{n} \cdot \sum_{i = 1}^{n} T_i
$$
is also a random variable, since it is a function of the samples $T_1, \dots, T_n$. This means that the mean could have been any value within the range of its probability distribution; in Figure \ref{fig:hyptest} we show two possible probability distributions for the sample mean of two populations of execution times. Since the distributions overlap, it could happen that we obtain two sample means $\overline{T}_1 < \overline{T}_2$ even though the expected value (the populational mean) is $E[T_1] > E[T_2]$. This is a major problem, as it allows a researcher to incorrectly conclude that their approach is better than the existing one. In order to mitigate this problem, one can make use of \textit{hypothesis testing}, which is further explained in the following.

\sigla*{CLT}{central limit theorem}
Testing hypotheses allows the researcher to better understand the probability that their conclusion was incorrectly inferred from the sample means obtained. This relies on the central limit theorem\footnotemark{} (CLT, see \citet{billingsley2008probability}), which tells us that for any random variable $X$, following \textit{any} probability distribution, then if we take samples $X_1, \dots, X_n$ and the sample mean $\overbar{X_n}$, we have that
\footnotetext{This theorem exists in more than one form; here we refer specifically to the ``classical'' version, in which $X_1, \dots, X_n$ are assumed to be independent and identically distributed. Later we might refer to Lyapunov's CLT, as it excludes the requirement of identical distributions.}
\begin{equation} \label{eq:central-limit}
    \lim_{n \rightarrow \infty} P\bigg[\: \frac{\overbar{X_n} - \mu}{\sigma / \sqrt{n}} \:\le\: x \:\bigg] = \phi(x)
\end{equation}
where $\mu$ and $\sigma$ are the populational mean and variance, respectively, and $\phi$ is the cumulative distribution of a standard normal distribution. The cumulative distribution uniquely characterizes a probability model, so we can conclude that as the sample size increases, the distribution of $\overbar{X_n}$ tends to $N(\mu, \sigma^2 / n)$ \footnote{Recall that if $X \sim N(\mu, \sigma^2)$ then $k X \sim N(k \mu, k^2 \sigma^2).$}, which is a gaussian whose variance also decreases as the sample size increase; thus, in Figure \ref{fig:hyptest}, the two histograms would get thinner if we increased the sample size, eventually eliminating the overlapping and, consequently, making the researcher more sure that the sample mean they got is near the true mean, with high probability. Testing hypotheses allows us to measure this degree of certainty.

The problem with (\ref{eq:central-limit}) is twofold. First, it only works when taking the limit $n \rightarrow \infty$, whereas researchers can never take sample of infinite size. Statisticians, however, have had to deal with such a problem for many decades, and experience made it well accepted that for a large class of random variables $n \ge 30$ already makes (\ref{eq:central-limit}) approximately valid \citep{walpole1993probability}. These random variables $X$ are expected to behave similarly to a normal, that is, they must be unimodal and reasonably symmetric. In this context, here we make the first approximation: we assume $\overline{X}$ has a normal distribution.

Under such assumption, we can calculate probabilities of interest from Figure \ref{fig:hyptest}. Consider that $\overbar{X_1}$ follows the blue histogram (to the left) and $\overbar{X_2}$ follows the other one, then we want to calculate $P(\overbar{X_1} < \overbar{X_2})$, which measures the probability of correctly asserting that the populational means follow $\mu_1 < \mu_2$; in contrast $P(\overbar{X_1} > \overbar{X_2})$ would measure the probability of saying that $\mu_1 > \mu_2$, when this is not true. We can rewrite
$$
P(\overbar{X_1} < \overbar{X_2}) = P(\overbar{X_1} - \overbar{X_2} < 0)
$$
and using known relations between linear combinations of normal distributions, we have
\begin{align} \label{eq:norm-diff-dist}
\overbar{X_1} - \overbar{X_2} \sim N(\mu_1 - \mu_2\;,\; \frac{\sigma_1^2}{n_1} + \frac{\sigma_2^2}{n_2})
\end{align}
from which we could easily calculate $P(\overbar{X_1} - \overbar{X_2} < 0)$, if it were not for the second problem that (\ref{eq:central-limit}) has.

\begin{figure}[htb]
    \centering
    \includegraphics[width=0.8\linewidth]{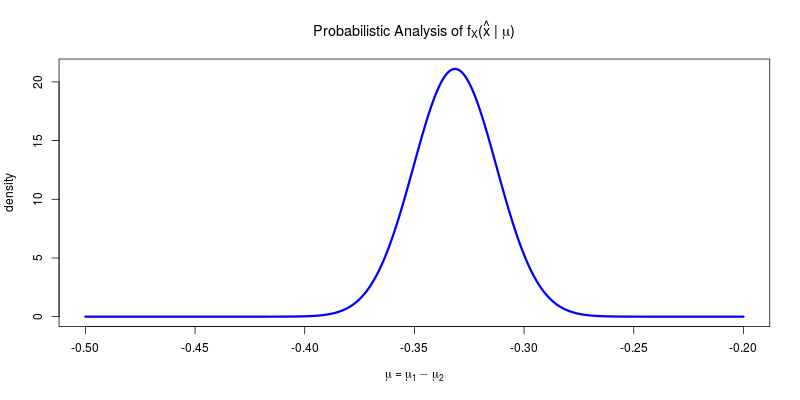}
    \caption[Conditional probability density of having sampled $\hat{x}$ from a normal distribution with given mean $\mu$ and variance.]{Conditional probability density of having sampled $\hat{x}$ from a normal distribution with given mean $\mu$ and variance as defined in the text. We emphasize that this graph does not represent a probability density because the variable on the x-axis ($\mu$) is not a random variable.}
    \label{fig:hyp_test_exploration}
\end{figure}

The second problem in (\ref{eq:central-limit}) is that $\mu = \mu_1 - \mu_2$ is unknown (as information on $\mu_1$ and $\mu_2$ is precisely what we are trying to infer from the sample), and $\sigma$ in most cases is also unknown. Suppose for now that $\sigma$ is known\rev{discuss sigma not known in an appendix?}. We have in hand a value $\hat{x}$ (commonly called a statistic), which was sampled from $Y = \overbar{X_1} - \overbar{X_2}$ that follows a normal distribution with parameters as in (\ref{eq:norm-diff-dist}). We do not know the parameter $\mu = \mu_1 - \mu_2$, but we can study the conditional probability density $f$ of having sampled $\hat{x}$ from the distribution given $\mu$; that is, study $f(\hat{x} \mid \mu)$. Suppose that we got $\hat{x} = -0.3315$\footnote{The values here come from a synthetic problem, in which two samples of size 100 were generated, one from a normal $N(1, 1)$, the other from $N(0.7, 1).$}, and that we know that
$$
    \frac{\sigma_1^2}{n_1} + \frac{\sigma_2^2}{n_2} = 0.02
$$
then the aforementioned conditional probability is shown in Figure \ref{fig:hyp_test_exploration}. This figure shows densities, so it should be taken with caution as it does not represent a probability. It is an approximation of the probability $P(\hat{x} - \epsilon < Y < \hat{x} + \epsilon)$ for very small $\epsilon$. Despite this, we can still draw conclusions.
For example, we see that if the mean was $\mu \ge 0$, meaning that algorithm 1 (here associated with $\mu_1$) is not faster than algorithm 2, it would be very unlikely to have sampled $\hat{x} = -0.3315$; in particular, the probability of having sampled any value below $-0.047$ is at most $1\%$, where $1\%$ happens when the mean of the normal is taken to be $0$, as higher values would yield even lower probabilities. In these circumstances, we can at least have high certainty that $\mu \ge 0$ (as initially assumed) is not true, meaning that $\mu_1 < \mu_2$ as desired.


What was done above is precisely what is usually called a hypothesis test, although using a less recipe-based, algorithmic process. In the usual process, we would have formulated the problem as two hypotheses
$$
    \begin{aligned}
        H_0&: \mu \ge 0 \\
        H_1&: \mu < 0
    \end{aligned}
$$
where $H_0$ is the null hypothesis that should be chosen as the one we have to assume true until we prove otherwise. In the case of algorithm comparison, it is better to assume the proposed algorithm is worse than the existing one, and then prove this wrong. Then we define a significance level $\alpha$, which above had been taken as $0.01$ (or $1\%$), which will be our criterion for rejecting the validity of $H_0$. This is done in the following way: 1) there exists a value $x_\alpha$ such that $\alpha = \sup_{\mu \in H_0} P(X < x_\alpha \mid \mu)$, where the supremum is not necessary as we know the worst case happens at $\mu = 0$; 2) then there is a probability of $1\%$ of having sampled any value below $x_\alpha$; 3) finally, if our sampled value $\hat{x} < x_\alpha$, we reject $H_0$, thus accept that our algorithm is faster. As a reference for other computer scientists, we highly recommend \citet{walpole1993probability} for the more algorithmic process of testing hypotheses; \citet{degroot2012probability} offers a more theoretical perspective that promotes a better interpretation of the process, and allows one to even modify tests to suit their problem better.

To perform the test, it was assumed that the sample mean $\overline{X}$ follows approximately a normal distribution, for the number $n$ of samples used. If the underlying distribution is ``well-behaved'', then it is a reasonable assumption for any $n \ge 30$ \citep{walpole1993probability}. In this context, do execution times follow a well-behaved distribution? What would be a reasonable $n$ to choose? We see (although very rarely) hypothesis tests being used in the literature -- as done by \citet{kadioglu2011algorithm} and \citet{saldanha2019high} for execution times, and by \citet{jindal2017reducing}, \citet{hassan2005comparison} and \citet{wang2015knowledge} for other measures concerning algorithms --, but no comments on the underlying distributions are made. The present monograph is thus an attempt to support hypothesis tests for comparing execution times of algorithms, provide means to better interpret the outcomes of these tests, and also shed light on the decision of the sample size $n$.

\rev{People usually do not do hypothesis tests. Which guarantees can we give to the reader, if they are provided with the mean and variances of the experiments? What about if only the minimum and maximum are given (assume a worst case scenario of a uniform distribution, for example; can we prove the uniform is the worst case?); the variance of the normal decreases with the sample size, so we could use Chebyshev's inequality to conclude something (There is also an interesting bound on sample variance, here: \url{https://en.wikipedia.org/wiki/Popoviciu\%27s_inequality_on_variances})?}

\section{Methods for Generating New Probability Distributions}
\label{sec:generating-new-distributions}

In this project we ended up using some distributions that are actually ``compositions'' of existing distributions. This section aims to explain such composition process, which will sate eventual curiosity, help interpreting how these distributions differ from other simple ones, and help understanding the naming logic of some distributions used here.

One way of composing distributions is as follows. Suppose we have a cumulative distribution function (cdf) $F(x)$, which we will call \textit{parent distribution} and whose support\footnote{The support is the domain of the cumulative/probability density functions (e.g., $\mathbb{R}$ ; $[0, 1]$ ; $[0, \infty)$).} is arbitrary; also consider a cdf $G(x)$ with support $[0, 1]$. Every cdf $H(x)$ follows the properties \citep{degroot2012probability}:
\begin{enumerate}[nosep,noitemsep]
    \item $\lim_{x \rightarrow \infty} H(x) = 1$;
    \item $\lim_{x \rightarrow -\infty} H(x) = 0$;
    \item $H(x)$ is non-decreasing function;
    \item $H(x)$ is right-continuous.
\end{enumerate}
The aforementioned $F$ and $G$ thus have the following signatures (the intervals on the codomains can be open or not):
\begin{align} \label{eq:F-G-definitions}
    F : \mathscr{E} \subset \mathbb{R} \rightarrow [0, 1] \hspace{1.5cm} G : [0, 1] \rightarrow [0, 1]
\end{align}
Which can be composed as $G \circ F : \mathscr{E} \rightarrow [0, 1]$, and what is interesting is that $G \circ F$ is most often also a cdf. Note that Property 3 holds because the composition of two non-decreasing functions is non-decreasing; that is, since $F'(x) \ge 0$ and $G'(x) \ge 0$:
\begin{align*}
    \frac{d}{dx} G(F(x)) = G'(F(x)) F'(x) \ge 0
\end{align*}
Property 4 also holds, as the composition of two right-continuous functions is right-continuous. Let $\lim_{x \rightarrow x_0^+} F(x) = L$ and $\lim_{x \rightarrow L^+} G(x) = M$, then we have:
\begin{align*}
    \lim_{x \rightarrow x_0^+} G(F(x)) = \lim_{x \rightarrow L^+} G(x) = M
\end{align*}
so $G(F(X))$ is right-continuous. Properties 1 and 2 can be proved, but it is somewhat more difficult to do so. The biggest problem is that the following equality (note that $\lim_{x \rightarrow \infty} F(x) = 1$):
\begin{align*}
    \lim_{x \rightarrow \infty} G(F(x)) = \lim_{x \rightarrow 1^-} G(x)
\end{align*}
only holds if $G$ is left-continuous at $x = 1$, which is not the case for many discrete distributions, but always holds for continuous ones. We thus conclude that for all continuous $G(x)$ and $F(x)$, $H(x) = G(F(x))$ will also be a cdf. Let us see how this works in practice.

\begin{figure}[H]
    \centering
    \includegraphics[width=\linewidth]{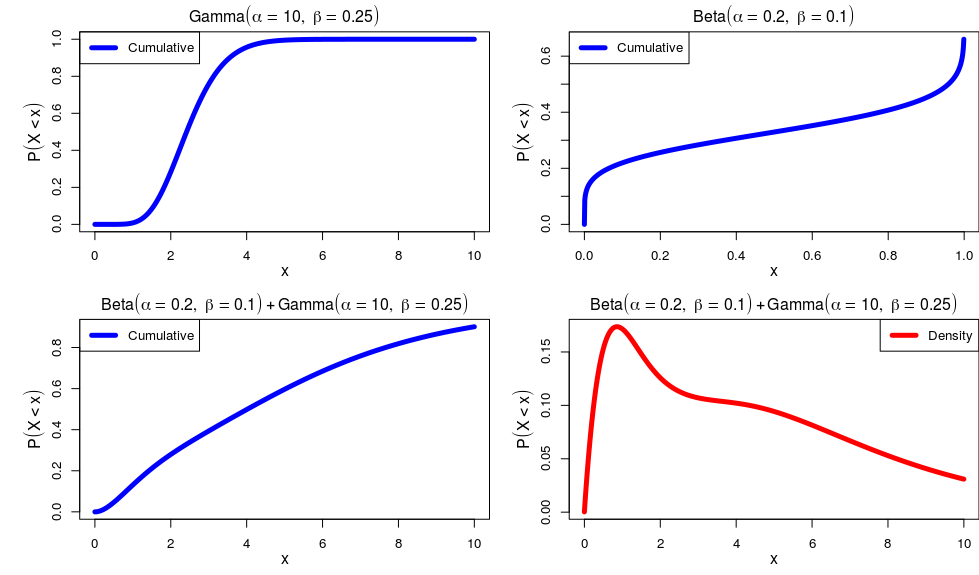}
    \caption[Example of a composition of Gamma and Beta distributions.]{Example of composing a Gamma parent distribution with a Beta distribution. From the top-left in a clockwise direction: 1) cdf of the Gamma used, 2) cdf of the Beta used, 3) cdf of the composition and 4) the pdf of the composition.}
    \label{fig:gamma-beta-composition}
\end{figure}

Consider $F$ is the cdf for a Gamma distribution with parameters $\alpha_1$ and $\beta_1$, which has support $[0, \infty]$; and $G$ is the cdf of a Beta distribution with parameters $\alpha_2$ and $\beta_2$, with support $[0, 1]$. Then applying the technique explained above, we can create a new cdf $G(F(x))$, resulting in a brand new distribution family with 4 parameters $(\alpha_1, \beta_1, \alpha_2, \beta_2)$, and its \sigla{pdf}{probability density function} can be found by differentiating the produced cdf. Figure \ref{fig:gamma-beta-composition} illustrates the composition process for parameters $(10, 0.25, 0.2, 0.1)$.

The process of building new distribution families using similar techniques is a very explored topic in Statistics \citep{cordeiro2011new,bourguignon2014weibull,torabi2014logistic,cordeiro2017generalized}. The process we described above was done in a similar way with the Kumaraswamy distribution, which has support in $[0,1]$, resulting in the Kw-CWG distribution we have used in this project \citep{afify2017new}. The same applies to the OLL-GG \citep{prataviera2017odd} distribution we also used.

\chapter{The Problem of Unknown Populational Minimum}
\label{chapter:inf-location-parameter}

\begin{figure*}[htb]
    \centering
    \includegraphics[width=0.8\textwidth]{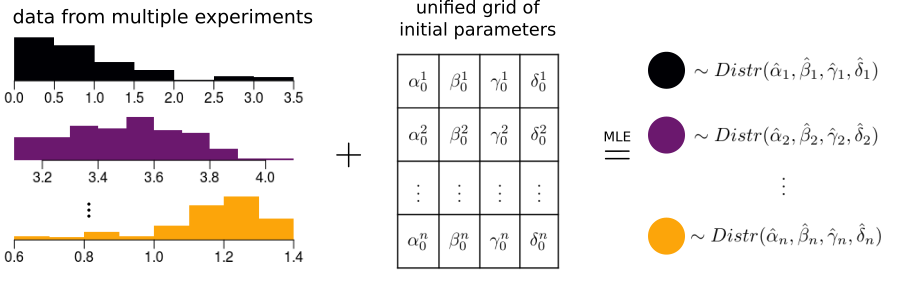}
    \caption[Main scenario to which we aim to contribute to.]{Main scenario to which we aim to contribute to. The experimenter has collected data from a number of different phenomena whose underlying probability distribution is believed to belong to a certain family $\mathscr{D}(\alpha, \beta, \gamma, \delta)$. We then would like to infer $\theta$ for each experiment. Usually, due to the variety of shapes and scales of the phenomena, one would have to define a grid of initial parameter for each phenomenon. We argue here that using our results one can define a single grid to perform inference for all the phenomena.}
    \label{fig:diagram-main-scenario}
\end{figure*}

During execution of this project, we faced a situation where multiple random variables (from different phenomena) are measured through experiments, and then a set of different distribution models, with a fixed set of initial parameters, are fitted to these variables (see Figure \ref{fig:diagram-main-scenario}). As discussed in Chapter \ref{chapter:unfolding}, this incurred multiple problems, mainly because performing maximum likelihood estimation upon the raw samples of execution times often did not converge.
In summary, we found it very fruitful to model the execution times as $T = \mathscr{m} + T'$, where $\mathscr{m}$ is the populational minimum and $T'$ the variable whose distribution is estimated. However, since $\mathscr{m}$ is unknown, it must be estimated, and this is a difficult problem.
In the following we further discuss the problem and propose some solutions to ease it in a general setting.


The aforementioned scenario, pictured in Figure \ref{fig:diagram-main-scenario}, can happen in various cases besides execution times. For example, the time between failures in a supply chain might follow an exponential with very low rate $\lambda \ll 1$, meaning the expected value of the minimum ($1 / (\lambda n)$) can be very high for a small sample of size $n$. Another example would be the time of a flight from Tokyo to Toronto, which clearly has a certain minimum value given by the limitations of airplane speed in the present age. These examples illustrate two cases that must be distinguished: one is when the underlying random variable has a long left tail, but its support remains being $[0, \infty)$; the other, when its support is $[a, \infty)$ for some $a > 0$.

In both these cases, it is most common to try to infer the underlying distributions using positive supported models (e.g., gamma, lognormal), maybe after subtracting the experimental data by a certain value $c$ that the statistician believes is the populational minimum of the underlying distribution. In either scenario, if the underlying distributions have left tails of various lengths, it becomes difficult to provide good initial conditions that are suitable to all these cases. Of course, simpler models can be given initial conditions based on method of moments, but the same cannot be said about more complex models such as generalized versions of Gamma and Weibull \citep{stacy1962generalization,mudholkar1993exponentiated}, nested models (e.g., Kumaraswamy- and Logistic-generalized distributions \citep{cordeiro2011new,torabi2014logistic}), mixture models, copula models \citep{jouini1996copula}, etc.

This is a dangerous situation when trying to seek the model that best fits the experimental data, because one often relies on the maximized likelihood for model selection, even when using information criteria \citep{anderson2004model};
because of that, it is a must that the maximized likelihood is indeed the maximum, which can be made impossible if good initial conditions are not given \citep{nocedal2006numerical}. This could in turn lead to biased conclusions in favor of the simpler models, which are less prone to optimization issues due to bad initial conditions. We therefore argue that the sample could be somehow modified in order to simplify the determination of good initial conditions. In the following we propose multiple methods, all of which aiming to:
\begin{itemize}[noitemsep,nosep]
    \item obtain higher overall maximized likelihood over multiple models;
    \item make it possible to recycle the same grid of initial parameters for performing inference over experiments from multiple phenomena; and
    \item minimize computational time spent in inference.
\end{itemize}



\section{Problem Formalization}
\label{sec:problem-formalization}

\begin{figure}[htb]
    \centering
    \includegraphics[width=0.6\linewidth]{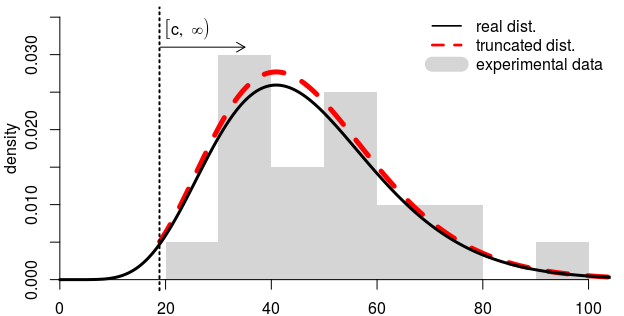}
    \caption[Comparison of the original distribution and its truncated version.]{Example of the first scenario analyzed. The underlying phenomenon is represented by a variable $X$ whose distribution is shown as the solid line. From such a distribution we take a sample (light grey histogram). Then we choose $c$ using methods to be discussed later, and fit a truncated distribution over $Y \approx X - c$ (dashed line).}
    \label{fig:formulation-fig1}
\end{figure}

First let $X$ be a random variable that follows a certain probability model with support $[0, \infty)$ \footnote{Note that the discussion presented here also applies to supports of type $[c, \infty),\, c \in \mathbb{R}$.}, and a sample $x_1, \dots,x_n$ taken from $X$. Consider the case where the experimental minimum is relatively high as illustrated in Figure \ref{fig:formulation-fig1}. By support, we mean the set in which the probability density is not zero, apart maybe from a subset of measure zero; hereafter, we consider all probability functions to be defined on the whole real line. In an attempt to reduce the space of initial conditions to explore, we model such variable as $X \approx c + Y$ with $Y \in [0, \infty)$ and $c \in \mathbb{R}_+$. Note that if this model was true, then the support of $X$ would be $[c, \infty)$, which is not true considering the initial assumptions. However, it seems reasonable to believe that if $P(X < c)$ is very low, then the loss incurred by such ``approximation'' would be negligible.

The approximation here consists of considering that the range of possible outcomes of $X$ begin at a certain $c$ that is not the true one.
We then would like to model the data under such a consideration; that is, find a model for $Y$.
If we have knowledge about the distribution family of $X$ and that its support begins at zero, then a good fit (asymptotically) would be achieved by selecting the distribution of $Y$ as being a truncated version of the distribution of $X$ (see Figure \ref{fig:formulation-fig1}), given by
\begin{equation*}
    f_{Y}(y \mid \theta) = \frac{f_X(y + c \mid \theta)}{1 - F_X(c)}, \; y \in [0, \infty)
\end{equation*}
where $f_Y,f_X$ are densities, $F_x$ is a cumulative distribution function (cdf), $\theta$ is a parameter vector and $c$ is given.
Notice that $f_Y(y) = \lambda f_X(y + c)$ for a constant $\lambda = 1 / (1 - F_X(c))$. Because of that, the likelihood over a sample $y_1, \dots, y_n$ is
\begin{align*}
    \mathcal{L}_{f_Y}(\theta \mid y_1, \dots, y_n) &= \prod_i f_Y(y_i \mid \theta) = \cdots
\end{align*}
and if $y_i > 0$ for all $i$, then
\begin{align*}
    \cdots = \prod_i \lambda f_X(y_i + c \mid \theta) &= \prod_i \lambda f_X(x_i \mid \theta) = \lambda^n \mathcal{L}_{f_X}(\theta)
\end{align*}
so that any $\theta$ maximizing the likelihood for $f_X$ will also maximize $\mathcal{L}_{f_Y}$, proving our initial statement that the best asymptotic fit would be a truncated version of $X$'s distribution. This happens regardless of $c$, so if we allowed $c$ to also be optimized, then it would be chosen to maximize $\lambda^n = 1 / (1 - F_X(c))$; from the monotonicity of $F_X$ maximization happens when $c$ approaches the sample minimum $\overbar{\mathscr{m}}$. We cannot have $c \ge \overbar{\mathscr{m}}$ because in such a case at least one of the $y_i$ would be negative, thus making $f_Y$ and the likelihood $\mathcal{L}_{f_Y}$ be zero.

The fact that $c$ has no influence in the best parameter $\theta$ found by \sigla{MLE}{Maximum Likelihood Estimation} is actually a problem here. Although truncation allows us to shift the support origin, it does not help us with our original objective of making the space of initial parameters easier to design.
This is related to the moments of $Y$ and $X$ being proportional to one another ($E[e^X] = e^c E[e^Y]$).


In the impossibility of using truncated models, we turn back to the original problem of finding a model for $X$ by modelling just $Y$. Recall that this means the support of $X$ is approximated as $[c, \infty)$, with $c$ being either fixed or given as a parameter of the distribution family.
Thus, consider that $X$ follows a certain distribution parametrized by $\theta_1 \in \Omega_1$, whereas the candidate distribution family that we use is parametrized by $(\theta_2, c) \in \Omega_2$ ($c$ can be fixed or not). We must then find the ``best'' $(\theta_2, c)$ in the parameter space. Let $x_1, \dots, x_n$ be a sample from the real distribution, then the average log likelihood can be expressed as (recall $c < \overbar{\mathscr{m}}$):
\begin{equation*}
    \frac{1}{n} \sum_{i = 1}^n \log f_2(x_i \mid \theta_2, c)
\end{equation*}
and as $n \rightarrow \infty$ we have, by the law of large numbers \citep{degroot2012probability}, its expected value:
\begin{equation} \label{eq:expected-likelihood}
    \int_0^\infty \log f_2(x_i \mid \theta_2, c) \,d F_1(x \mid \theta_1)
\end{equation}
which gives us a metric that we would like to maximize. With this we are seeking the model that obtains highest expected likelihood over data generated by the real underlying distribution $F_2$.

\begin{figure}[htb]
    \centering
    \includegraphics[width=0.6\linewidth]{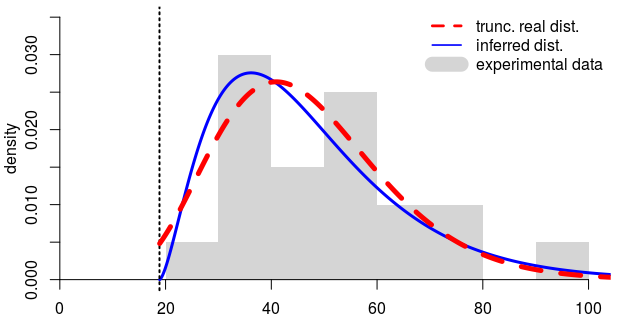}
    \caption[Comparison of the original distribution and the fit obtained by shifting its support.]{The ideal distribution would be the truncated version of the real underlying distribution (dashed line). However, if we exclude truncated distributions as argued in the text, we end up with a suboptimal solution (solid line), obtained here by maximizing the average likelihood shown in Eq. \ref{eq:expected-likelihood}.}
    \label{fig:formulation-fig2}
\end{figure}

\begin{figure}[htb]
    \centering
    \includegraphics[width=0.6\linewidth]{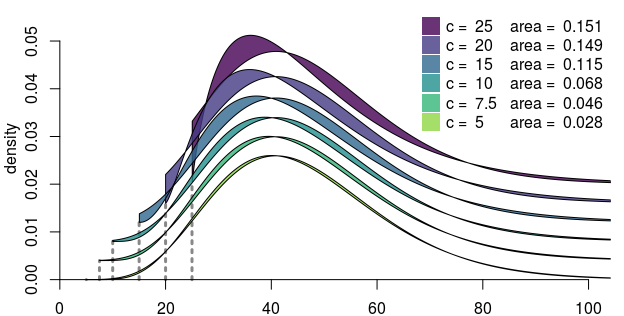}
    \caption[Difference between our fitted model and the ideal truncated version.]{Considering a certain Gamma distribution $\mathscr{D}$ with a long left tail, this figure shows the best Gamma approximations to the ideal truncated version of $\mathscr{D}$, when performing inference on support $[c, \infty)$. Each shaded area shows the region between two curves: i) the ideal truncated distribution, and ii) the curve obtained by maximizing the average log-likelihood. The curves have been displaced on the y-axis for better visualization.}
    \label{fig:areas-between-functions}
\end{figure}

We now consider the case where the distribution of $Y$ is inferred from the same distribution family of $X$. Here, the optimal solution (truncated version of $X$) is almost never included in the inference search space\footnote{In order for this to happen, the distribution would need to be memoryless, and we know only the exponential and geometric distributions are memoryless \citep{ross2010introduction}.}. Instead, the resulting distribution will be an approximation of this optimal solution, as shown in Figure \ref{fig:formulation-fig2}. The area between the curves is illustrative of the difference between their cumulative probabilities, so it can be used to have an idea of how much they differ. We had constrained $c$ to be lower than the sample minimum $\overbar{\mathscr{m}}$; for small samples, this leaves a large range over which $c$ could lie. Figure \ref{fig:areas-between-functions} shows what happens when we perform inference for different choices of $c$. High values, nearer the sample minimum, will result in more disparate distributions than the truncated one. On the other hand, low values that are nearer to the origin of the original variable $X$ (lower values would also work) tend to yield a distribution more similar to the ideal. We consequently face a tradeoff as high values of $c$ is what allows recycling a grid of initial values for various phenomena.

The above discussion has so far considered that the statistician knows that the underlying random variable is supported on $[0, \infty)$. However, it is often the case that this is not known with sufficient certainty. In fact, for distributions with long left tails, which are the main object of study here, we probably will not observe any values very near zero, even if the underlying distribution was indeed supported on $x \ge 0$.
Nevertheless, the case of support $[c, \infty)$ is also covered if the coordinate system is translated.



\section[Proposed Methods for Performing the Inference Procedure]{Proposed Methods for Performing the Inference \\ Procedure}
\label{sec:proposed-methods}


Due to the aforementioned hindrance in determining whether the underlying phenomenon is supported on $x \ge 0$, we argue that all semi-finite supported random variables must be considered as belonging to the interval $[\mathscr{m}, \infty)$ until proven otherwise. As it is an estimator of the populational minimum $\mathscr{m}$, any value of $c$ above the sample minimum $\overbar{\mathscr{m}} = \min \{ x_1, \dots, x_n \}$ does not make sense. In this scenario, given a sample $x_1, \dots, x_n$ we would like to find the underlying distribution within a parametrized family supported on $[c, \infty)$.

In order to ease the determination of initial parameters for the subsequent inference process, $c$ is to be chosen regardless of the real value of $\mathscr{m}$, merely aiming for having $P(X < c)$ be low enough and $c$ be as near the sample minimum as possible, since this will minimize the approximation losses discussed in Section \ref{sec:problem-formalization}. Our choices are then to either estimate $c$ and then perform inference over $Y = X - c$ using a family $\mathscr{D}(\theta)$, or to add $c$ as a location parameter of such a family, which then becomes $\mathscr{D}(\theta, c)$. Recall that we eliminated the possibility of truncation previously. We analyze both possibilities, and in the end propose a third alternative that deviates slightly from the usual procedure of classical inference. We remind that the objective is maximize likelihood and minimize computational cost.

\medskip
\textbf{I) Inferring the Location Parameter.} Let $X$ represent the underlying phenomenon with support $[\mathscr{m}, \infty)$. We want to model it using a family $\mathscr{D}(\theta)$ of distributions supported on $\mathbb{R}_+$, though shifted to $[c, \infty)$. That is, we actually model $Y$ such that:
\begin{align*}
    f_Y(y \mid \theta, c) = \begin{cases}
        f_x(y - c \mid \theta),\;&\text{ if } y \ge c \\
        0,\;&\text{otherwise}
    \end{cases}
\end{align*}
in which case, by abuse of notation, we say $Y \sim \mathscr{D}(\theta, c)$ with $c$ constrained to lie in the interval $[0, \overbar{\mathscr{m}})$ ($\overbar{\mathscr{m}}$ the sample minimum). With this, MLE can then be performed to find $\theta$ and $c$.




\medskip
\textbf{II) Estimating the Location Parameter.} Since $c$ is strongly related to the populational minimum $\mathscr{m}$, in that the value of $c$ that maximizes likelihood when $n \rightarrow \infty$ is $\mathscr{m}$, then it makes sense to estimate it from a sample. With the estimate $\hat{c}$ we can then perform inference using a positive supported distribution as usual after subtracting $\hat{c}$ from the sample (recall that $X = c + Y$). Taking the sample minimum to estimate it, besides being very biased, also frequently results in the likelihood term for the smallest sample item being zero, rendering MLE impossible. After subtracting the estimate from the sample $y_1, \dots, y_n$, the smallest one ends up being $y_i = x_i - \hat{c} = 0$; the problem here resides on the fact that many distributions require $f(0) = 0$ for a large range of their parameters, such as the Gamma distribution whenever the shape parameter is $\alpha \neq 1$.

Shifting that estimate slightly to the left is thus needed, maybe by multiplying it by some factor. But what should this factor be? In our experience, deciding this automatically to various data sets with different shapes and scales happened to be quite difficult. For example, taking $\hat{c}$ to be the upper end of the rescaled interval $0.95 \cdot [0, \overbar{\mathscr{m}})$ worked for data sets with smaller values, but not for larger ones where it was shifted too far from $\overbar{\mathscr{m}}$. In order to find better alternatives, we rely on order statistics.

The sample minimum $\overbar{\mathscr{m}}$ has a known cumulative distribution \citep{mood1950introduction}:
\begin{align} \label{eq:minimum-cdf}
    F_{\overbar{\mathscr{m}}}(x) = 1 - [ 1 - F_X(x) ]^n
\end{align}
for a sample of size $n$ of a variable with cdf $F_X(x)$.  Therefore we can expect the sample variance to have, for small samples, a variance similar to that of the underlying variable, and we can expect the variance to reduce as $n$ grows. This reasoning brings us to the first estimator:
\boldmath\begin{align}
    \hat{c}_1(x_1, \dots, x_n) = \overbar{\mathscr{m}} - \frac{|\overbar{\mathscr{m}} \cdot CV|}{\log_k(n)}
\end{align}\unboldmath
where $CV$ is the variation coefficient\footnote{$CV = \overbar{\sigma} / \overbar{\mu}$, sample variance divided by sample mean.} of the sample and $k$ is an arbitrary logarithm basis. Here and in subsequent estimators, we use the modulus because the underlying variable can have negative $\mathscr{m}$, in which case $\overbar{\mathscr{m}}$ can also be negative. The interpretation is that we are moving $\overbar{\mathscr{m}}$ to the left, with an intensity that is directly proportional to the data variability and inversely proportional to the sample size. Our experience showed $k = 10$ to be quite useful and computationally cheap. It is worth noting that taking the coefficient of variation eliminates, to a certain extent, problems caused by the scale of the data, since it involves a division by the sample mean. This estimator has the advantage of simplicity, and even though it gives a very rough estimator of the desired low quantile, it appears to work very well in practice.

We now turn to more complex alternatives, that have more theoretical grounds. Although there are many parametric approaches for estimating quantiles \citep{gardes2005estimating,de2018high}, estimating very low quantiles is a problem that has not yet been solved in a sufficiently ``general'' way. That is, most parametric solutions rely on assumptions about the underlying distribution or quantile functions (constraints on their derivatives, for example). To maintain generality (and because this later proved itself to work well), we opt for a general non\babelhyphen{hard}parametric approach, using the empirical cdf $F_n(x)$ over $n$ samples as main tool. Given a sample $x_1,\dots,x_n$, $F_n(x)$ is defined as the proportion of sample items lower than $x$, thus forming an increasing step function. Uniform convergence of $F_n(x)$ to $F(x)$ is given by the Glivenko-Cantelli theorem\footnote{This, as well as all other results used hereafter, require independent and identically distributed (iid) sampling.}, so for sufficiently large $n$ we have information about the probability $P(X \le \overbar{\mathscr{m}}) \approx F_n(\overbar{\mathscr{m}}) = 1/n$ of a next sample to be lower than the current sample minimum. As this number decreases, the less we can expect the populational minimum to be lower than the actual sample minimum, meaning that we can then define a second estimator:
\boldmath\begin{align} \label{eq:estimator-cv}
    \hat{c}_2(x_1, \dots, x_n) = \overbar{\mathscr{m}} - \frac{\hat{\sigma}}{n}
\end{align}\unboldmath
where we also embody the hope that the deviation $\mathscr{m} - \overbar{\mathscr{m}}$ is proportional to the sample standard deviation. A tighter estimate follows by noticing that the Law of Iterated Logarithm \citep{vapnik1998statistical} gives the rate of convergence:
\begin{align*}
    \lim_{n \rightarrow \infty} \; \sup_{l > n} \; \forall x \, \big| F(x) - F_l(x) \big| \le \sqrt{\frac{\ln \ln l}{2 l}}
\end{align*}
Now considering that $F_n(\overbar{\mathscr{m}} - \epsilon)$ is zero for any $\epsilon > 0$, we must have for sufficiently large $n$:
\begin{align*}
    F(\overbar{\mathscr{m}} - \epsilon) \le F_n(\overbar{\mathscr{m}} - \epsilon) + \sqrt{\frac{\ln \ln n}{2 n}} = \sqrt{\frac{\ln \ln n}{2 n}}
\end{align*}
which can then substitute the $1/n$ in Eq. (\ref{eq:estimator-cv}):
\boldmath\begin{align} \label{eq:estimator-iter-logarithm}
    \hat{c}_3(x_1, \dots, x_n) = \overbar{\mathscr{m}} - \hat{\sigma} \cdot \sqrt{\frac{\ln \ln n}{2 n}}
\end{align}\unboldmath

\begin{figure}[htb]
    \centering
    \includegraphics[width=0.6\linewidth]{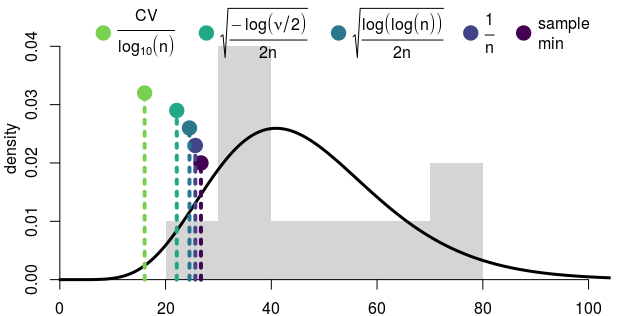}
    \caption{The low quantile estimators based on the data represented by the histogram in light gray. The data was generated from the density shown as a black line.}
    \label{fig:estimators-example}
\end{figure}

The Dvoretzky–Kiefer–Wolfowitz inequality \citep{massart1990tight} can also be invoked, which provides a different way to view the estimator. The inequality is:
\begin{align*}
    P( \sqrt{n} \sup_x | F_n(x) - F(x) | > \lambda ) \le 2 \exp(-2 \lambda^2)
\end{align*}
and by doing the necessary manipulations, we derive that the following will hold with probability $1 - \nu$:
\begin{align*}
    \sup_x | F_n(x) - F(x) | \le \sqrt{\frac{-\ln(\nu / 2)}{2n}}
\end{align*}
so if we choose $\nu$ to be very low, we can expect $F(\overbar{\mathscr{m}} - \epsilon)$ to be lower than or equal to the right-side of the above equation. Following the same logic as previously, we define another estimator:
\boldmath\begin{align} \label{eq:estimator-dvoretzky}
    \hat{c}_4(x_1, \dots, x_n) = \overbar{\mathscr{m}} - \hat{\sigma} \cdot \sqrt{\frac{-\ln(\nu / 2)}{2n}}
\end{align}\unboldmath
which offers a probabilistic view, instead of the previous asymptotic view given by the Law of Iterated Logarithm. Figure \ref{fig:estimators-example} illustrates all of these estimators.


\medskip
\textbf{III) Iterative Determination of the Location Parameter.} Inference by MLE begins with the assumption that the underlying distribution comes from a certain family. Under this assumption, we do have a lot of information about the underlying cdf and pdf. We intend to use this information to our advantage here.

The cdf of the sample minimum is given by Eq. (\ref{eq:minimum-cdf}). By inverting this equation we obtain the quantile function of the minimum:
\begin{align} \label{eq:quantile-of-minimum}
    F_{\overbar{\mathscr{m}}}^{-1}(q \mid \theta) = F_X^{-1}(1 - (1 - q)^{1/n} \mid \theta)
\end{align}
With this, the median of the sample minimum is given by $F_{\overbar{\mathscr{m}}}^{-1}(q \mid \theta)$, under the assumptions that the underlying distribution resides in the specified family and has parameter $\theta$. The median can be seen as a good ``guess'' for what the sample minimum should be, and so the sample should be shifted so that the sample minimum coincides with such a guess. That is, we want to find $c$ such that $\overbar{\mathscr{m}} - c = F_{\overbar{\mathscr{m}}}^{-1}(0.5 \mid \theta)$, and the function that undergoes MLE is
\begin{align*}
    f_Y(y \mid \theta) = f_X(y - \overbar{\mathscr{m}} + F_{\overbar{\mathscr{m}}}^{-1}(0.5 \mid \theta) \mid \theta)
\end{align*}
However, it is very unusual to place a sample measurement (in this case $\overbar{\mathscr{m}}$) in the density function. This method has the advantage of not increasing the number of parameters, which favors better values of information criteria and reduce chance of overfitting. One worry here is that the aforementioned modification to the density function might make the likelihood surface very irregular and thus harder to optimize over. Despite that, experiments have shown it works quite well, as we show in the following; theoretical analysis of the optimization surface will be done in the future.

\section{Experimental Results}
\label{sec:min-estimator-experiments}

\begin{figure}[htb]
    \centering
    \includegraphics[width=0.49\linewidth]{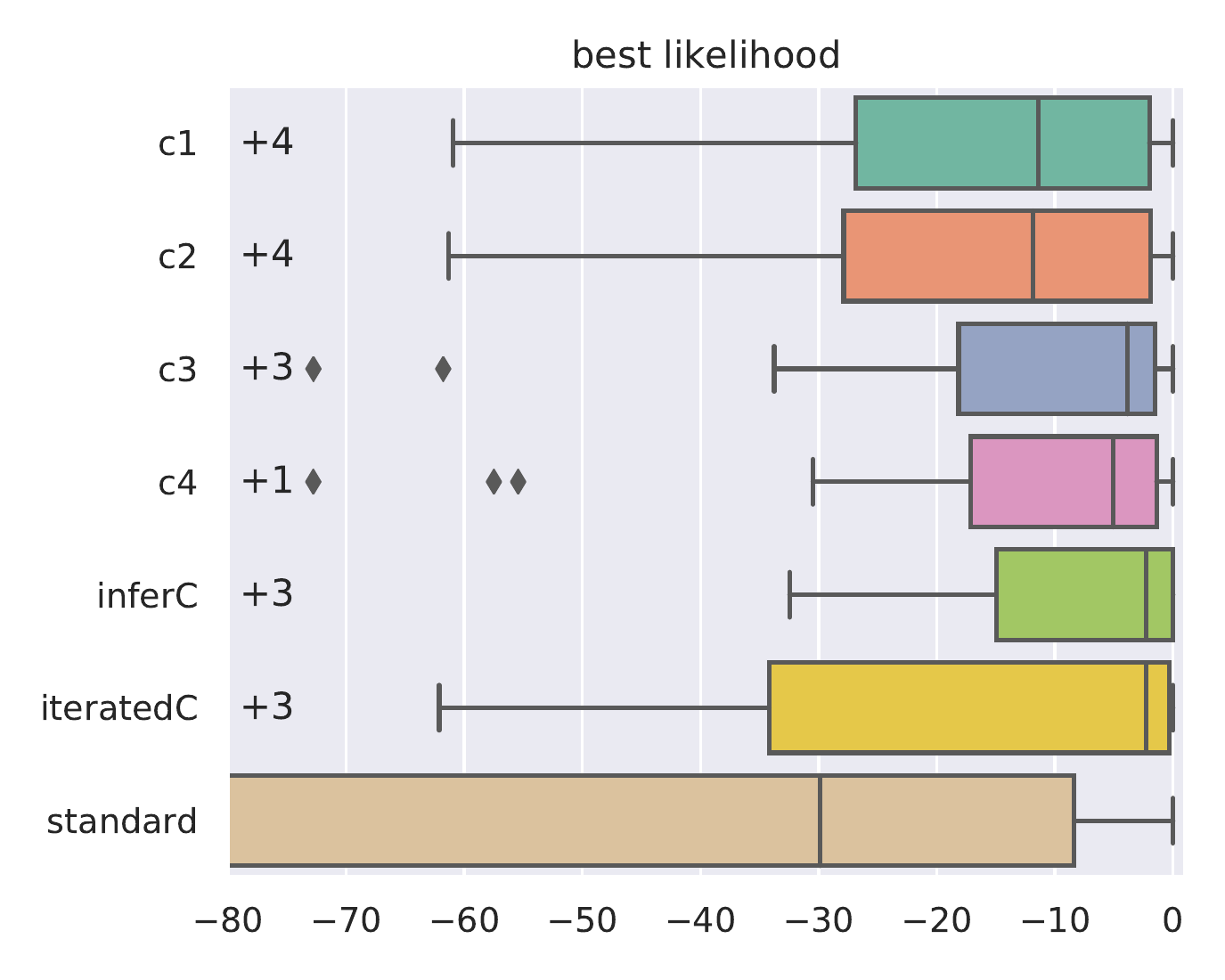}
    \includegraphics[width=0.49\linewidth]{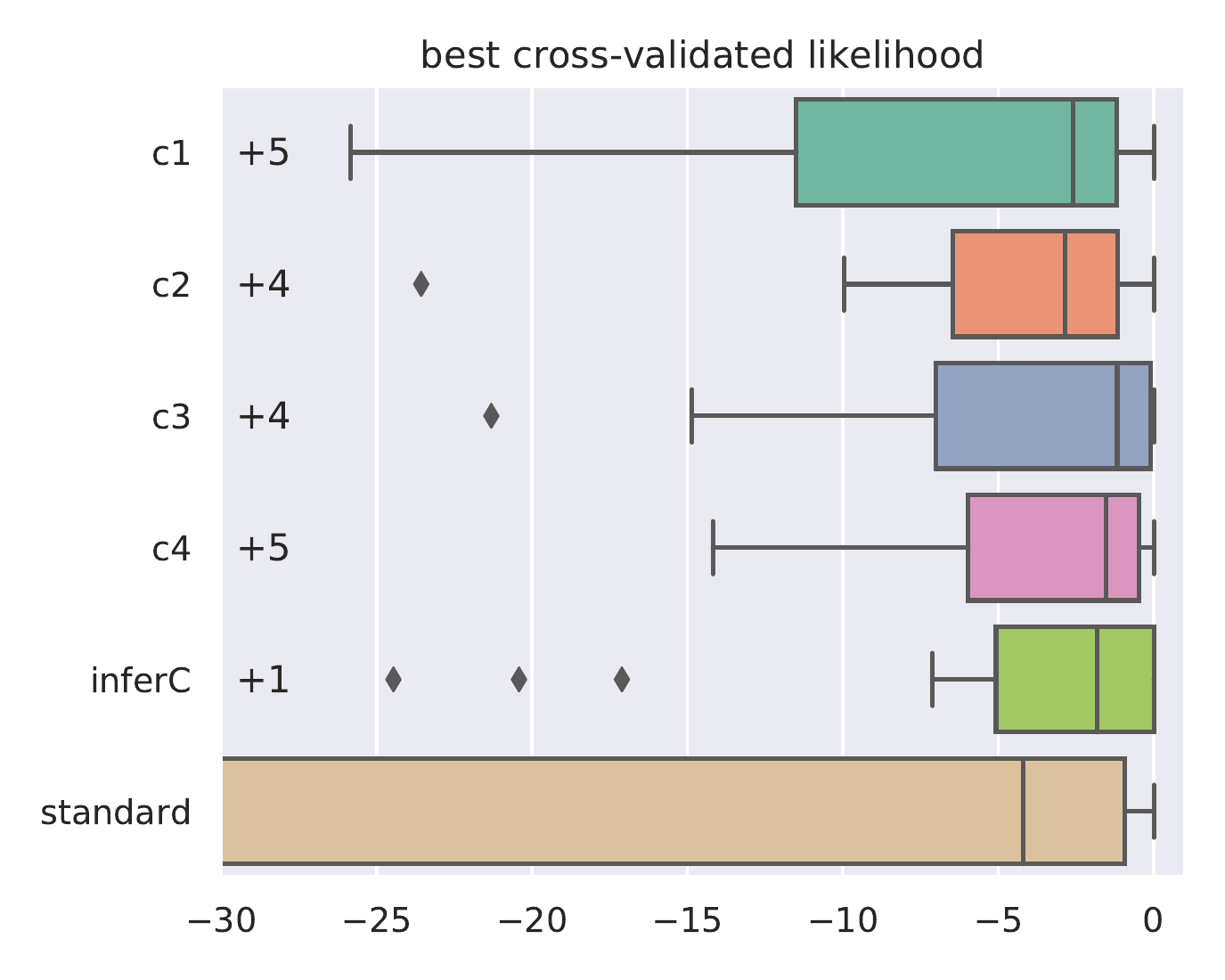}
    \caption[Comparison of inference methods based on the best likelihood they yielded for each of the 37 sample sets.]{Comparison of inference methods based on the best likelihood (left) and best cross-validated likelihood (right) they yielded for each of the 37 sample sets. A value of zero means a case where that inference method achieved the best result among all other methods; negative values shows a disparity from the best method.}
    \label{fig:boxplots-inf-method-quality}
\end{figure}

So far we have tested the proposed methods only with the execution times data obtained during this project. We had multiple sets of execution time samples, and performed MLE using multiple distribution families. For each sample set and inference method, one of these distributions yields the best likelihood $-2\hat{l}$, which we take as being the measure of quality of that inference method. Finally, for each sample set one of the inference methods has the best quality, so we subtract the quality of each inference method by that best value. This results in $37$ quality measures for each inference method, which are plotted as boxplots in Figure \ref{fig:boxplots-inf-method-quality}. Here, a value of zero means it was the best inference method in that sample set, and negative values shows its deviation from the best method.

\rev{Could talk about why it is valid to take the difference between likelihoods.}

First consider the boxplot concerning the likelihood $-2\hat{l}$ only. In this case the figure indicates that the \texttt{inferC} method (adds $c$ as a parameter of the distribution) has the best median as well as a low dispersion, so overall it performed best. However, this increases the parameter count, which in general is not desirable. Among the models that do not increase parameter count, estimators $\hat{c}_3$ and $\hat{c}_4$ displayed very similar performances, comparable to that of \texttt{inferC} and much better than the standard inference method. The main problem with the standard method is what we mentioned in the beginning of the chapter: the optimization algorithm fails to find the point of maximum likelihood, consequently giving some sort of preference to the simpler distribution families. The proposed \texttt{iteratedC} had a median very close to the best one, though it has a longer tail towards worse likelihood values. We consider its performance really good, especially considering that it has a solid theoretical reasoning (together with \texttt{inferC}), so we believe it should not be discarded. Rather than that, this method has a parameter $q$ that we fixed as $0.5$ in Section \ref{sec:proposed-methods}, so we hypothesize that this could be tuned to yield even better results.

There are a few notable changes when turning to the cross-validated metric. First of all there is a remarkable improvement in the performance of all $c_k$ estimators relative to \texttt{inferC}, with $c_3$ achieving the best median. \texttt{inferC} still displayed a very good performance, proving that it is quite robust, more than one could expect from a method that increases the parameter count and, as a consequence, the chance of overfitting. In general, all of our proposed methods (except \texttt{iteratedC}) proved to be robust, in the sense that their performance do not change significantly when cross validating. \texttt{iteratedC} displayed terrible performance when cross validating, but we attribute this more to how the method was formulated than to its actual usefulness. We intend to further investigate this in the future, in order to improve the method formulation.

Minimizing computational time of the inference process is also one of our objectives. For each sample set and each distribution model, each inference method took some amount of time to be performed. Recall that MLE optimization is done with many initial parameters (that we defined), which can lead the optimizer to converge or not. Furthermore, optimization is faster if the initial parameters are nearer to the optimal value, so using only one value as reference would probably be too biased towards the choice of initial parameters. We thus consider: 1) the average time elapsed in each of these optimizations, i.e. the time taken per initial parameter by the inference method in question; and 2) the average time for all initial parameters that led the optimizer to converge.

\begin{figure}[htb]
    \centering
    \includegraphics[width=0.8\linewidth]{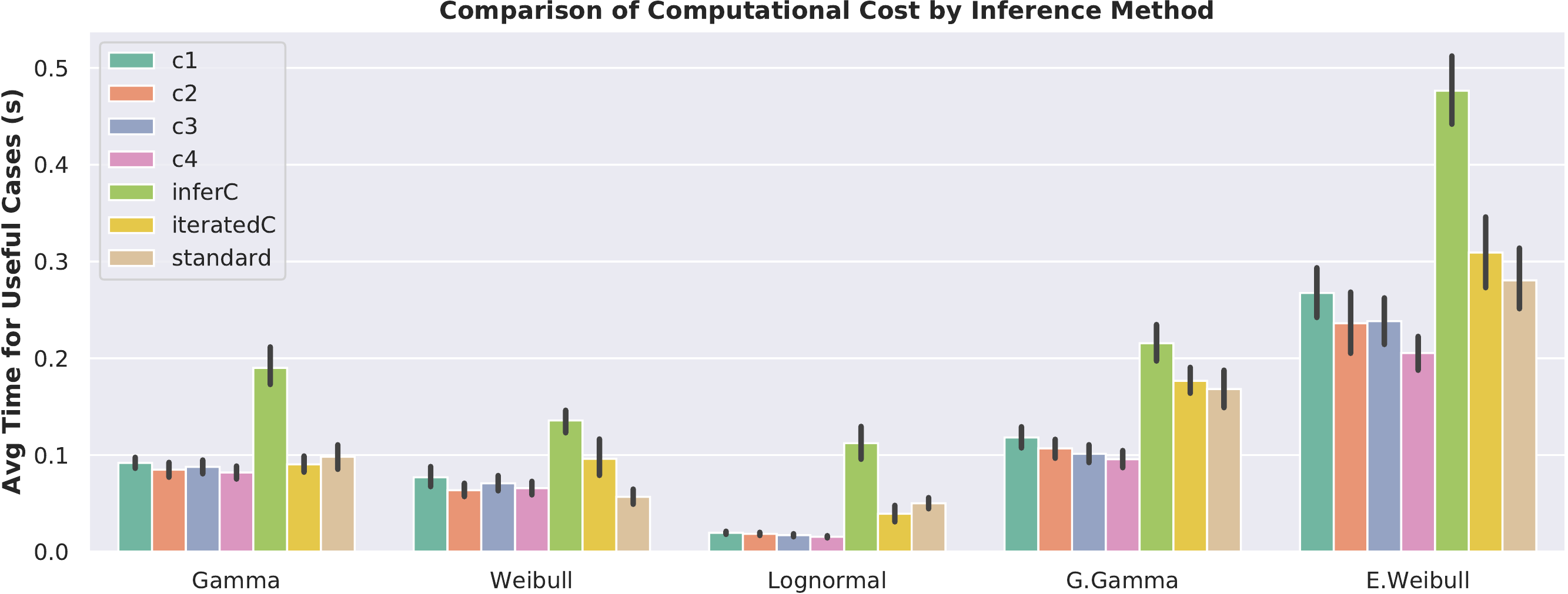}\\[0.4em]
    \includegraphics[width=0.8\linewidth]{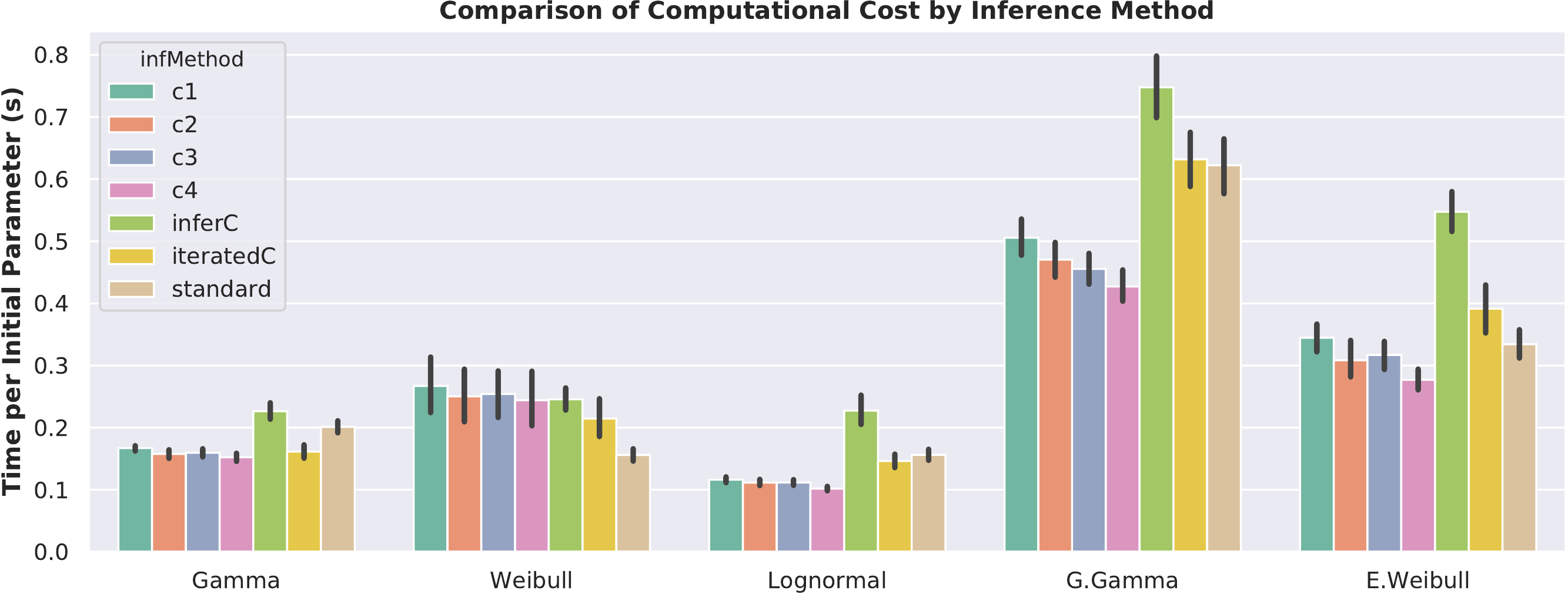}
    \caption[Comparison of computational cost of the inference methods for each distribution family.]{Comparison of computational cost of the inference methods for each distribution family, considering only the initial parameters that led to convergence (top), and all initial parameters (bottom). The error bars show a $95\%$ confidence interval, obtained through bootstrapping.}
    \label{fig:barplot-infmethod-per-initial-parameter}
\end{figure}

Figure \ref{fig:barplot-infmethod-per-initial-parameter} shows these average times for each distribution family\footnote{Experiments performed in the Euler supercomputer \href{http://www.cemeai.icmc.usp.br/Euler/index.html}{www.cemeai.icmc.usp.br/Euler/index.html} using an Intel Xeon E5-2680v2 2.8 GHz and 128 GB of RAM.}. As expected, \texttt{inferC} was the slowest in most cases, as it has one extra degree of freedom to optimize; it is followed by \texttt{iteratedC} (probably because the optimization surface became more irregular, as discussed previously) and the standard method. Interestingly, the $c_k$ estimators were almost always faster than the standard method, with $c_4$ being the fastest one. This proves that the $c_k$ estimators we proposed do not only help achieving better likelihood values, but also hasten the optimization process. In most cases the time difference was not too big; in the worst case (lognormal), the $c_k$ estimators were ${\sim}10$ times faster than \texttt{inferC}, and ${\sim}4$ times faster than the standard method. Nevertheless, even small differences can make a huge difference when there is a large number of distributions, initial parameters and sample sets to undergo distribution fitting. If the whole inference process takes 5 minutes, a difference of only $20\%$ in time would make it 4 minutes only. The $c_k$ estimators suffer some degradation when considering all initial parameters, i.e. including those that did not converge; this shows that non-convergent cases can take more time than convergent ones. Despite that, the whole discussion above holds, except for the Weibull distribution, which we will consider as an anomalous case here, and not discuss it further.

To sum up, the inference methods $\hat{c}_3$, $\hat{c}_4$ and \texttt{inferC} yield very good likelihood values, and all of them were robust. $\hat{c}_3$ and $\hat{c}_4$ were slightly more robust, and managed to decrease the time taken to perform MLE, most of the times even faster than the standard method. On the other hand \texttt{inferC} was the slowest method of all tested.

\chapter{Determining the Distribution of Execution Times}
\label{chapter:unfolding}

This project consisted of understanding the problem domain (which comprises most aspects that influence the execution time of a program), modelling the problem probabilistically, executing experiments and, finally, analysing the data obtained experimentally. This chapter is dedicated to expose details of such course of investigation.

\section{On the Suitability of Using Probability Models}
\label{sec:suitability-of-prob-models}

\begin{figure}[htb]
    \centering
    \includegraphics[width=0.48\linewidth]{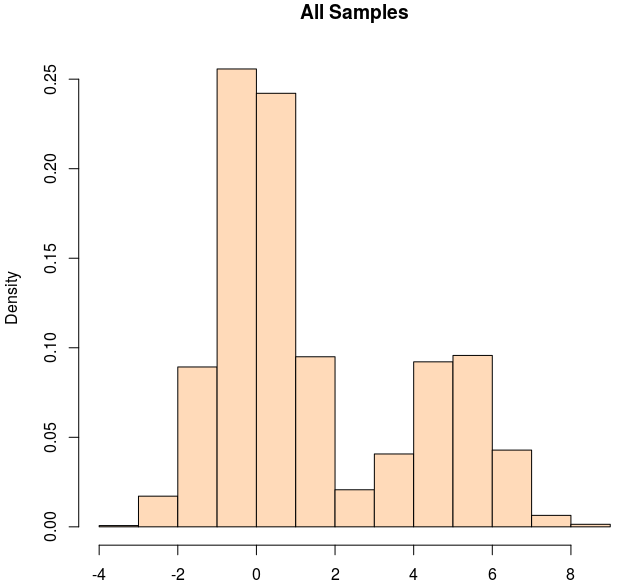}
    \includegraphics[width=0.48\linewidth]{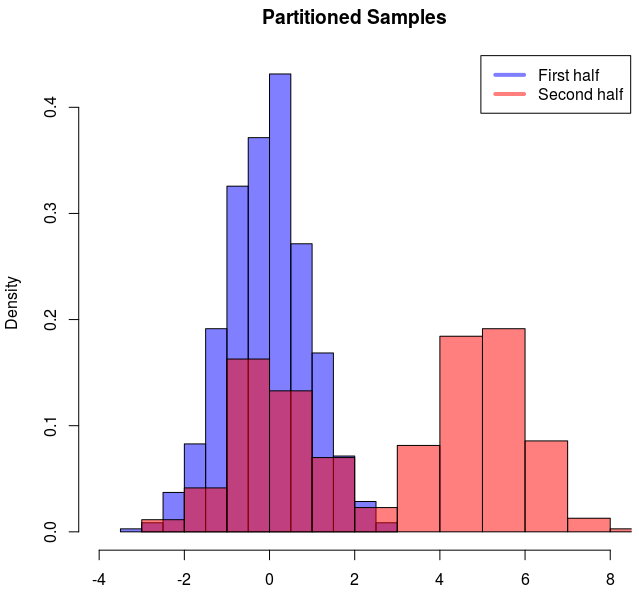}
    \caption[Histograms displaying a change of the underlying distribution at some point in time.]{Possible histogram for execution times of a program (to the left), and the histograms of the two halves of the whole sample (to the right), showing that the underlying population suffered a permanent change at some point in time.}
    \label{fig:possible-histogram-problem}
\end{figure}

The randomness present in execution times of programs is obvious. Despite this, the correct way of modelling it probabilistically is not as clear; in fact, before modelling any problem probabilistically, some time must be spent in interpreting the underlying phenomenon being observed and foreseeing what could go wrong in the model. We illustrate one possible problem in Figure \ref{fig:possible-histogram-problem}. Suppose that the program under study (henceforth referred to as the \textit{experimental program}) is run a number $N$ of times and the execution times are recorded, then this might yield a histogram such as that on the left of the figure. One could naively accept such histogram as representative of the \textit{whole population} of execution times that the program could have yielded; however, knowledge about how the computer works internally forces us to be more skeptic about it.

If we plot the histograms for first and second halves of the $N$ samples, we might get what is shown to the right of Figure \ref{fig:possible-histogram-problem}, which clearly shows that something is wrong, at least for sufficiently large $N$ (e.g., $N > 200$), as there is the theoretically justified expectation that histograms of each half to be very similar to the histogram for the whole sample. More specifically, something has changed at some point in time, so the underlying populational distribution is different for the first half than for the second one. It could have happened, for example, that during the experiment some daemon process (e.g., an automatic backup began) started running and competing with our experimental program for the computer resources \citep{tanenbaum2015modern}, which led to degradation of the execution times. This can be dealt with in a few different forms:
\begin{itemize}[itemsep=0.3em, parsep=0.3em]
    \item we can run the experimental program every some large interval $T$ of time, for a total period of a few days, so that the histogram reflects the probability distribution averaged over all possible states of the machine (e.g., busy or idle);
    \item we can try to make sure the machine will be sufficiently stable during experiments;
    \item the problem can be modelled as a time series, which would allow for a better prediction of the next execution time, based on previous ones \citep{box2015time};
    \item the samples of the experiment could be investigated for anomalies such as the one described above, using techniques from the area of novelty detection \citep{markou2003novelty}.
\end{itemize}
Deciding what to do and how to do, this is the problem of \textit{modelling} and requires a deeper understanding of the problem domain (in our case, computer programs), which we try to better formulate in the following.

\subsection{The Problem of Concurrent Processes}
\label{sec:problem-concurrent-processes}

A program is a binary file containing machine instructions, each of which is a number of bits describing what the computer CPU must do. A \textit{process} is a program in execution, that is, it has been loaded into primary memory and its instructions can be read at any time (controlled by the operating system) by the CPU for executing them \citep{tanenbaum2015modern}. During execution, the process can use all the computer resources, namely the CPU cores and cache memories, primary memory, secondary memory (e.g., hard disk), coprocessors such as GPUs, and so on \citep{patterson2013computer,tanenbaum2016structured,stallings2003computer}. The time to perform any operation with such resources is random, so they play a major role in the randomness of the program's total execution time.

If there are multiple processes running in the same system, they compete for the aforementioned resources and will degrade the execution time of our experimental program. Modern CPUs are able to execute multiple processes at the same time; each can be in their own CPU core, in which case the different processes will share: cache memories (level 1 cache is usually local to the core, so no sharing happens), primary memory and access to the system bus (and consequently, access to disk, GPUs, etc). It is also possible for two processes to run in the same core (i.e., hyperthreading), in which case they will also share the core pipeline and the lowest level caches (advanced modern CPU technologies are well covered in the work of \citet{hennessy2011computer}). Up to this point, each process runs continuously. If there are too many processes actively running, then in most operating systems processes will alternate between ``executing'' and ``suspended'' states, where the latter means the process is not running at all \citep{tanenbaum2015modern}. Finally, a last layer of randomness is introduced in cloud computing because programs are often run inside virtual machines, which adds a layer of abstraction between the program and the computer resources.


In the context of cloud computing, which is the focus of this monograph, machines tend to always have some process in active state. At each time point, all existing active programs will degrade the execution time of the experimental program in $T_D$ units of time, where $T_D$ is a random variable. This $T_D$ depends on the number of active processes and their types, that is, whether they are limited by CPU, memory or Input-Output operations \citep{tanenbaum2015modern}. In practice, it would be difficult for the cloud provider to keep track of the type of each application, so we disregard it here. Ideally, the information of how many active processes would be available, in which case it seems reasonable, based on what was discussed above, to model the execution time for the experimental program as
$$
    T = T_P + T_D(n)
$$
where $T_P$ is the probability distribution for the time of the program in a machine with no other active processes, $n$ is the number of active processes, and $T_D(n)$ is the distribution of the time degradation caused by these active processes, and this distribution depends on $n$. By combining regression analysis and statistical inference \citep{degroot2012probability}, this model already allows us to estimate a distribution for $T$.

\rev{One could argue that the probability distribution for each instruction could be estimated, and the distribution for the program would be the sum of these distributions. This is false.}

\rev{If the system has the stability described above, we could model $T_D(n)$ averaged over all possible $n$, that is $f(x) =(1/N) \sum_n f_{T_D}(x \mid n)$ or even estimate a probability distribution for $n$.}

\subsection{The Problem of Modelling the Program's Inputs}
\label{sec:problem-program-inputs}

A second problem arises when considering that a program can be given different input data. How does this affect the execution time? In order to understand this, we must introduce some definitions. Every program can be described by a control flow graph (CFG) $G = (\mathcal{N}, \mathcal{E})$ \citep{allen1970control}, in which nodes $\mathcal{N} = ( n_1, n_2, \dots )$ represent a linear sequence of instructions, and edges $(n_i, n_j, p) \in \mathcal{E}$ represent jumps that make the CPU stop reading instructions from the current node $n_i$ and begin reading the ones in the target node $n_j$. These jumps are most often conditional branches that decide the target node based on whether a certain predicate $p$ (such as \texttt{i < 10}) is true of false, which is indicated as labels on edges of the graph. A graph $G$ with $k$ nodes $n_1, \dots, n_k$ always has an entry node in which computation begins, and one or more exit nodes that terminate the program; for simplicity, and without loss of generality\footnote{Note that if there are multiple exit nodes, we can turn them into normal nodes and make them point to a new, unified exit node.}, we will assume a single exit node, $n_1$ always being the entry point and $n_k$ always the exit one. Each execution of the program corresponds to a walk through this graph beginning on $n_1$ and ending on $n_k$; an execution path is thus a sequence of nodes $(n_1, a_1, a_2, \dots, a_m, n_k)$, with $a_i \in \mathcal{N}$ and $m \in \mathbb{Z}_+$, visited during an execution.

\begin{figure}[htb]
    \centering
    \includegraphics[width=0.7\linewidth]{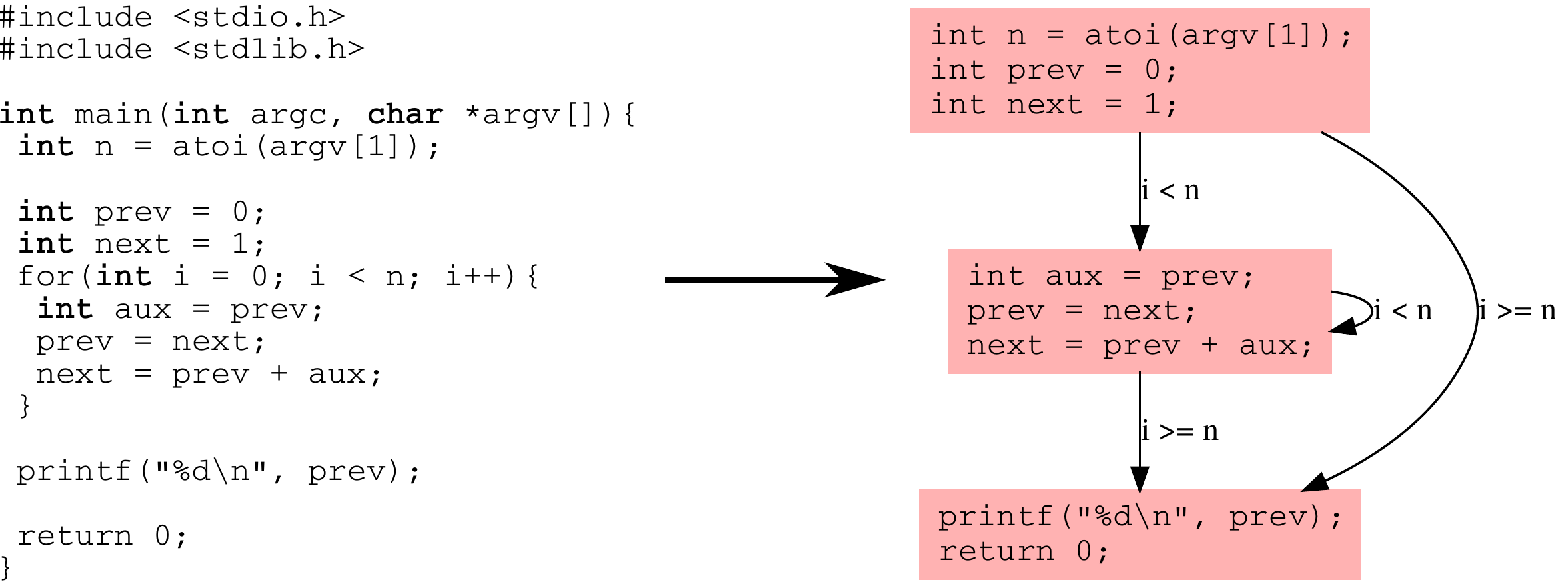}
    \caption{Source code for calculating the $n$-th fibonacci number (left) and its associated control flow graph (right).}
    \label{fig:cfg_img}
\end{figure}

An example is illustrated in Figure \ref{fig:cfg_img}, where a program that computes the $n$-th fibonacci number is converted to its associated CFG with nodes $n_1, n_2, n_3$ numbered from top to bottom following our previous assumptions. The input of this program is $n$, the index of the fibonacci number that should be computed. As each instruction of the program operates deterministically upon the results of previous operations, we can conclude that the program itself is deterministic, that is, for a specific input $n_0$ the execution path of the program is uniquely defined.

The problem introduced by the variety of possible inputs is that the mix of instructions executed by the program can change a lot between two different inputs. The execution time probability distribution for the program depends a lot on this mix of instructions, so the variety of inputs must somehow be accounted for. One way to do this is to treat the input as a random variable, and assign a probability for the program to receive any given input, as done by \citet{braams2016deriving}. Other ways specific to the WCET problem were discussed in Section \ref{sec:wcet}.

A third way to deal with inputs, which will be the approach adopted here, is to assume the program will always follow the same path or a set of paths that is sufficiently similar to model their execution times under the same probability distribution. This assumption is reasonable in at least the following two scenarios:
\begin{itemize}[nosep]
    \item the program will be run multiple times without changing the input, which is common for scientists that are performing experiments with their novel algorithms;
    \item the program will be run multiple times with different inputs, but the input variable that will be changed is not expected to change the execution path significantly.
\end{itemize}
As an example of the latter scenario, take the problem of protein structure prediction, which is a hard problem that received a lot of attention as to how it can be distributed over multiple machines in order to improve speed \citep{alva2016mpi} or throughput \citep{kim2004protein}; consequently, it is a strong candidate for being executed on the cloud \citep{kajan2013cloud,mrozek2015scaling,yachdav2014predictprotein}. There exist a number of stochastic optimization algorithms designed to solve this problem, such as done in the works of \citet{brasil2013multiobjective}, \citet{gao2018incorporation} and \citet{cutello2005multi}, and one of their characteristic is the \textit{number of iterations} parameter defined by the user, which controls how long the algorithm should be run for. The execution path of these algorithms in general depend more on such parameter than on the input protein itself, so running the program on different proteins using the same number of iterations tends to yield sufficiently similar execution paths. Note that cloud applications like this one, which are mostly run by researchers and scientists of varied knowledge areas, go under the special name of \textit{scientific workflows} \citep{shishido2018genetic} and, as a consequence of what was just discussed, is a major focus of the present work.

It is worth noticing that some algorithms, such as the stochastic optimization we mentioned above, make use of random number generators. In this case, each instruction is no longer deterministic, as we required earlier. This is fine as long as the generated numbers do not change the predicates along the execution path, which could, for instance, change the number of iterations in a for-loop and thus greatly impact the execution time. This often does not happen in most evolutionary algorithms, such as ant colony optimization, particle swarm optimization and genetic algorithms \citep{yu2010introduction}. Therefore, we can still regard execution paths as similar in this context.

\rev{Reference for the 90/10 rule \citep{dixit1991spec}}


\section[Selection of Distributions and Experimental Programs]{Selection of Distributions and Experimental \\ Programs}
\label{sec:selection-dist-and-programs}

\sigla*{Kw-CWG}{Kumaraswamy complementary Weibull geometric}
There exist a huge number of probability distribution families that could be used here to fit the execution times of a program in an idle system (with no other active processes), obtained through experiments. Due to the nature of the problem in hand, it is more adequate to choose distributions whose support is $\mathbb{R}_+$ (also called lifetime or survival models \citep{klein2006survival}), as execution times are naturally continuous and positive. This restricted class of distributions is still enormous, so we decided to begin by using two large models with many parameters, which consequently are very flexible; after fitting the data, it could provide information on which other distributions to choose. The chosen large models are the Kumaraswamy complementary Weibull geometric (Kw-CWG, \citet{afify2017new}) which has 5 parameters $\alpha \in (0, 1)$ and $\beta, \gamma, a, b > 0$, and the odd log-logistic generalized gamma (OLL-GG, \citet{prataviera2017odd}) with parameters $\alpha, \tau, k, \lambda$. The density functions of Kw-CWG and OLL-GG are respectively:
\begin{align*}
    f(x) &= \alpha^a \beta \gamma a b (\gamma x)^{\beta - 1} \exp[-(\gamma x)^\beta] \cdot
    \frac{\{1 - \exp[-(\gamma x)^\beta]\}^{a-1}}{\{ \alpha + (1 - \alpha) \exp[-(\gamma x)^\beta] \}^{a+1}} \cdot \\[0.5em]
    &\text{\hspace{15em}} \cdot \bigg\{ 1 - \frac{\alpha^a[1 - \exp[-(\gamma x)^\beta]]^a}{\{ \alpha + (1 - \alpha) \exp[-(\gamma x)^\beta] \}^a} \bigg\}^{b - 1} \\[0.8em]
    f(x) &= \frac{
        \lambda \tau (x / \alpha)^{\tau k - 1} \exp[ - (x/\alpha)^\tau]\{ \gamma_1(k, (t/\alpha)^\tau )[1 - \gamma_1(k, (t/\alpha)^\tau)] \}^{\lambda - 1}
    }{
        \alpha \Gamma(k) \{ \gamma_1^\lambda(k, (x/\alpha)^\tau) + [1 - \gamma_1(k, (x/\alpha)^\tau)]^\lambda ] \}^2
    }
\end{align*}
where $\Gamma(x)$ is the gamma function defined as $\Gamma(k) = \int_0^\infty x^{k-1} e^{-x} dx$ which has the property $\Gamma(n) = (n-1)!$ when $n \in \mathbb{N} \setminus \{ 0 \}$, so it should be viewed as an extension of the factorial to the real (or complex) domain \citep{boyce2012elementary}; and $\gamma_1(k, z)$ is the incomplete gamma ratio defined as $\gamma_1(k, z) = \int_0^z x^{k-1} e^{-x} dx / \Gamma(k)$, where the integral is very similar to the one in $\Gamma(k)$, with just a different interval of integration. The Kw-CWG model is a generalization of the complementary Weibull geometric distribution \citep{louzada2011complementary}, and OLL-GG results from composing a generalized gamma with an odd log-logistic distribution \citep{cordeiro2017generalized}; both comprise (for fixed values of its parameters) a number of other distributions as sub-models.

Initial experiments indicated that the underlying distribution of execution times are mostly unimodal and two-tailed; there were some exceptions where two modes could be observed. In this light, we immediately exclude distributions such as the exponential, which is one-tailed. Due to their wide usage, the gamma, Weibull and lognormal models were used; much of their possible shapes are unimodal and two-tailed as desired. Their generalized versions, the generalized gamma (by \citet{stacy1962generalization}) and the exponentiated Weibull (by \citet{mudholkar1993exponentiated}), were also considered in hope that they would fit the data better. Also, since the normal distribution is broadly used in literature, it is also used; however, the fact that it is a distribution over the whole real line $\mathbb{R}$ could place it in some sort of unfair disadvantage relative to the others. To avoid this problem, we also used a truncated form of the normal distribution, that is, if $f$ is the density function of the normal, then we take $g(x) = f(x) / \int_{x > 0} f(x) dx$ (with $x > 0$) where the denominator is a normalization factor to force $g(x)$ integrate to 1, as all probability distributions should. The probability distribution functions for these distributions are shown in Table \ref{tab:density-functions} where $x > 0$ unless specified otherwise and $\Phi$ is the cumulative distribution of the usual normal distribution.

\begin{table}[htb]
    \centering
    \footnotesize
    \begin{align*}
        \text{Weibull: }&\;\; f(x) = \frac{k}{\lambda} \bigg( \frac{x}{\lambda}\bigg)^{k-1} \exp[- (x/\lambda)^k] &\lambda, k > 0 \\[0.5em]
        \text{Gamma: }&\;\; f(x) = \frac{x^{\alpha-1} \exp[-(x / \theta)]}{\theta^\alpha \Gamma(\alpha)} &\alpha, \theta > 0 \\[0.5em]
        \text{Gen. Gamma: }&\;\; f(x) = \frac{b x^{bk-1} \exp[-(x/a)^b]}{a^{bk} \Gamma(k)} &a, b, k > 0 \\[0.5em]
        \text{Exp. Weibull: }&\;\; f(x) = \frac{\sigma \nu x^{\sigma - 1}}{\mu^\sigma} \frac{\exp[-(x/\mu)^\sigma]}{(1 - \exp[-(x/\mu)^\sigma])^{1 - \nu}} &\sigma, \nu, \mu > 0 \\[0.5em]
        \text{Normal: }&\;\; f(x) = \frac{1}{\sqrt{2 \pi \sigma^2}} \exp\bigg[\frac{-(x-\mu)^2}{2\sigma^2}\bigg] &x \in \mathbb{R}, \mu \in \mathbb{R}, \sigma > 0\\[0.5em]
        \text{Trunc. Normal: }&\;\; f(x) = \frac{1}{\sqrt{2 \pi \sigma^2}} \exp\bigg[\frac{-(x-\mu)^2}{2\sigma^2}\bigg] \frac{1}{1 - \Phi(0 \mid \mu, \sigma)} &\mu \in \mathbb{R}, \sigma > 0 \\[0.5em]
        \text{Lognormal: }&\;\; f(x) = \frac{1}{x \sqrt{2 \pi \sigma^2}} \exp\bigg[\frac{-(\ln x-\mu)^2}{2\sigma^2}\bigg] &\mu \in \mathbb{R}, \sigma > 0
    \end{align*}
    \caption{Density functions of the probability distributions used.}
    \label{tab:density-functions}
\end{table}

Seven models were considered for the random variable $T$, which represents execution times of a program in an idle system. Initially $T$ was considered as having its own probability distribution, which was estimated from the data. As this proved itself problematic, we used the other six methods shown in Chapter \ref{chapter:inf-location-parameter} for dealing with location parameters.

In order to test the hypothesis that execution times of programs can be modelled using the same parametrized distribution, we need a representative set of programs and computer platforms on which these programs will be run. In an attempt to be representative, we have chosen three programs that exercise different components of the computer: one CPU-bound program, one memory-bound and one IO-bound (more specifically, bounded by disk usage) \citep{tanenbaum2015modern}. Respectively, the programs\footnote{\label{footnote:project-page}These programs and other code resources can be found in \href{https://github.com/matheushjs/ElfProbTET}{github.com/matheushjs/ElfProbTET} or \href{https://mjsaldanha.com/sci-projects/3-prob-exec-times-1/}{mjsaldanha.com/sci-projects/3-prob-exec-times-1/}.} are: 1) calculation of the Mandelbrot set\footnote{This set involves repeated calculations of the function $f: \mathbb{C}^2 \rightarrow \mathbb{C}$, $(z, c) \mapsto z^2 + c$. It is fractal as it exhibits complex structures in arbitrarily small open sets (see \citet{alligood1996chaos})}, 2) calculation of the shortest path in a random graph using Dijkstra's algorithm and 3) performing random insertions in a table of a database. Concerning machines, we attempted to build a selection as diverse as possible in terms of hardware, within the limits of infrastructural resources available to us; Table \ref{tab:machines} shows this selection. We highlight that the selection includes CPUs of two different vendors, as well as HD and SSD disks, which we expect to be among the most influential factors on the probability distribution

Concerning machines, we attempted to build a selection as diverse as possible in terms of hardware, within the limits of infrastructural resources available to us; Table \ref{tab:machines} shows this selection. We highlight that the selection includes CPUs of two different vendors, as well as HD and SSD disks, which we expect to be among the most influential factors on the probability distribution of execution times. Unfortunately, all CPUs follow the same \sigla{ISA}{instruction set architecture} specification (x86-64, also known as AMD64), which we believe is a limitation of this work; future work should include other ISA specifications (e.g., PowerPC, ARM).


\begin{table}[htb]
    \centering
    \footnotesize
    \newcommand{\celltext}[1]{\small #1}
    \begin{tabular}{@{}lm{10em}m{12em}m{10em}@{}} \toprule
        \textbf{Machine Name} & \textbf{CPU} & \textbf{Memory} & \textbf{Disk} \\\midrule
        \celltext{Andromeda} & \celltext{AMD FX(tm) 8350 Eight Core Processor} & \celltext{4x Corsair 8GB DIMM DDR3 Synchronous 800 MHz} & \celltext{Seagate HD 2TB S\-T\-2\-0\-0\-0\-D\-M\-0\-0\-1\-1\-E\-R\-1} \\\midrule
        \celltext{HalleyHD}  & \celltext{Intel Core i7 4790 CPU 3.60GHz} & \celltext{4x AMI 8GB DIMM DDR3 Synchronous 1600 MHz} & \celltext{Seagate HD 2TB S\-T\-2\-0\-0\-0\-D\-M\-0\-0\-1\-1\-C\-H\-1} \\\midrule
        \celltext{HalleySSD} & \celltext{Intel Core i7 4790 CPU 3.60GHz} & \celltext{4x AMI 8GB DIMM DDR3 Synchronous 1600 MHz} & \celltext{Kingston SSD 240GB \-S\-A\-4\-0\-0\-S\-3} \\\midrule
        \celltext{Helix}     & \celltext{Intel Core i5 4440 CPU 3.10GHz} & \celltext{4x Kingston 4GB DIMM DDR3 Synchronous 1333 MHz} & \celltext{Kingston SSD 240GB \-S\-A\-4\-0\-0\-S\-3} \\\bottomrule
    \end{tabular}
    \caption[Machines used for performing experiments.]{Machines used for performing experiments. Most of them are from the Laboratory of Distributed Systems and Concurrent Programming \url{http://lasdpc.icmc.usp.br/}.}
    \label{tab:machines}
\end{table}

\section{Experiments and Results}
\label{sec:experiments-results}

\begin{figure}[htb]
    \centering
    \includegraphics[width=\linewidth]{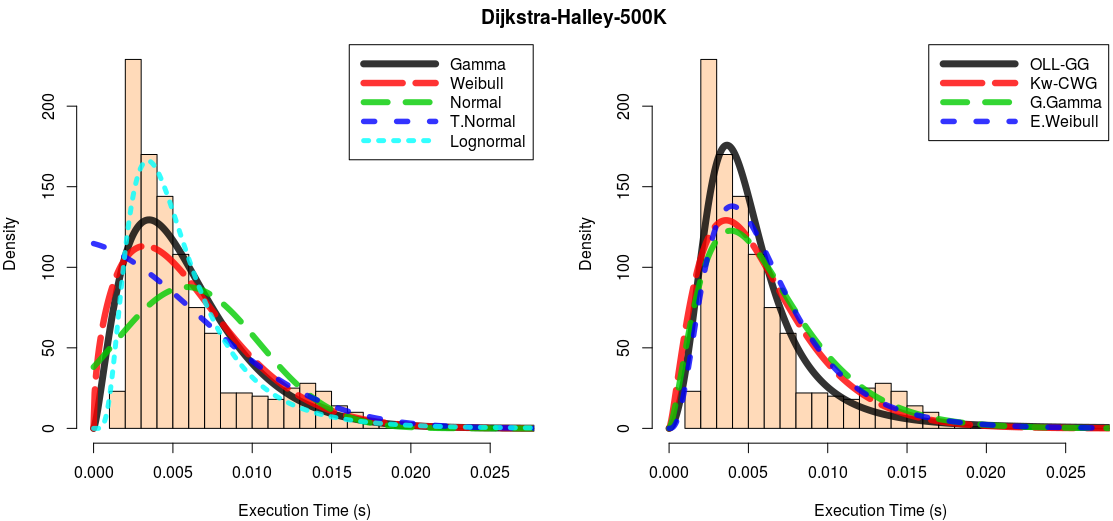}
    \caption[Example of result obtained by performing maximum likelihood estimation on a sample of execution times.]{Example of result obtained by performing maximum likelihood estimation on a sample of execution times. Each line reflects the best parameters found for the specified distribution family.}
    \label{fig:example-mle-fitting}
\end{figure}

Each of the three selected programs were executed 1000 times in each of the four selected machines for each different input chosen\footnote{There were 3 different inputs for the mandelbrot program, and 4 for the dijkstra and disk database programs.}, yielding 1000 samples of execution times for each triple program-input-machine. Then each of these sets of 1000 samples underwent maximum likelihood estimation to find best-fit parameters for each of the selected distribution families (an example is given in Figure \ref{fig:example-mle-fitting}), considering each of the seven proposed probabilistic modellings for execution times (see Section \ref{sec:selection-dist-and-programs} for details on such decisions). The experimental programs were implemented in C and Python, and the statistical analysis was made in the R language; the reader can find all code and experimental data in our public repository or the homepage of this project\textsuperscript{\ref{footnote:project-page}}.

A few technical details are worth mentioning. First, each program was run on an almost-idle machine, that is, no user application was being run, but system software and other daemons that occasionally become active were. Also, each set of 1000 samples was visually checked over whether the histograms of the first and second halves differed too much\footnote{Experimental data is also available in this project's home page; see footnote \textsuperscript{\ref{footnote:project-page}}.}, which would characterize a lack of identical distribution, as discussed in Section \ref{sec:suitability-of-prob-models}. Third, care was taken so that random generators that could change the execution path of the programs (see Section \ref{sec:problem-program-inputs}) received a fixed seed. Finally, the Generalized Gamma, Kw-CWG and OLL-GG probability distributions were not readily available in R, so we implemented it ourselves. Since the Kw-CWG implementation in R was slow (due to it being an interpreted language), we made an efficient C++ implementation, which uses parallel computing if supported by the user's machine. All these distributions were published it in the official R repository\footnote{See \href{https://cran.r-project.org/package=ggamma}{cran.r-project.org/package=ggamma}, \href{https://cran.r-project.org/package=elfDistr}{cran.r-project.org/package=elfDistr} and \href{https://cran.r-project.org/package=ollggamma}{cran.r-project.org/package=ollggamma}.}.

In order to evaluate how well the distributions fit the data, we used well-known metrics specific to this problem. Let $\hat{l}$ be the logarithm of the maximum likelihood obtained, the metrics used are
\begin{center}
    \vspace{-0.5em}
    \begin{tabular}{l l}
        $AIC = -2\hat{l} + 2k$ & $CAIC = -2\hat{l} + 2kn/(n-k-1)$ \\
        $HQIC = -2\hat{l} + 2k\log(\log(n))$ & $BIC = -2\hat{l} + k\log(n)$
    \end{tabular}
\end{center}
where $k$ is the number of parameters of the distribution and $n$ is the sample size.
These metrics are known as Akaike (AIC), consistent Akaike (CAIC), Hannan-Quinn (HQIC) and Bayesian (BIC) information criteria \citep{anderson2004model}, each of which try to penalize models with too many parameters to prevent overfitting; the maximized likelihood $\hat{l}$ was also used in the usual form\footnote{This form is merely so that it is directly comparable to the information criteria values.} $-2\hat{l}$. Finally, we also 5-fold cross validated the $-2\hat{l}$ metric, by fitting the models in subsamples of size $800$ and then calculating the metric in the remaining $200$\footnote{Cross validation is also a way to control overfitting, but the underlying theory (discussed by \citet{vapnik1998statistical} and \citet{devroye1996probabilistic}, for example) is completely different than in information criteria.}. For all metrics, the lower its value the better the model fits the data.
Besides these, we also recorded the time elapsed for performing the optimization procedure, as estimation should not take too long if used in a practical situation of stochastic scheduling (see Section \ref{sec:cloud-computing}).

\begin{figure}[tb]
    \centering
    \includegraphics[width=\linewidth]{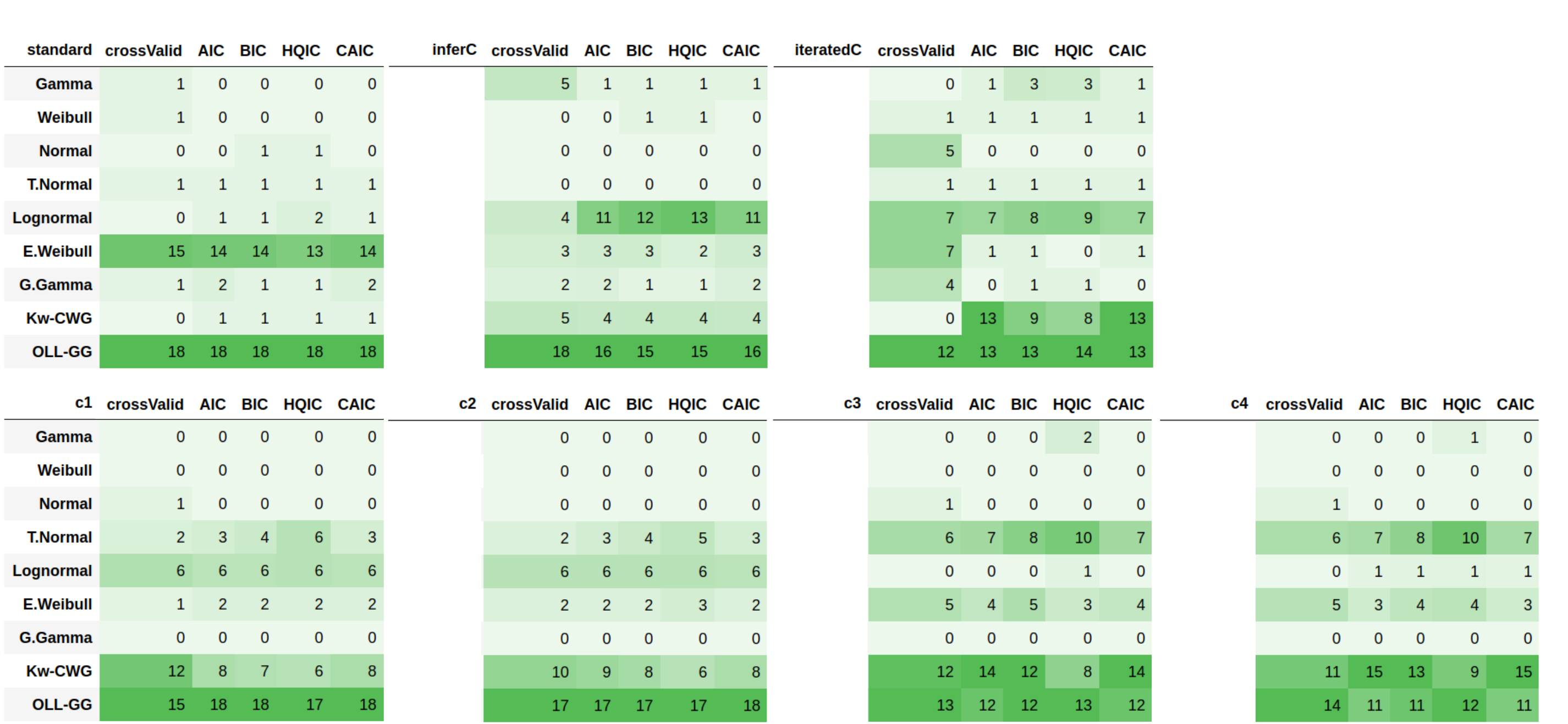}
    \caption[Number of times each distribution model achieved the best goodness-of-fit for each modelling for the execution time random variable.]{Number of times each distribution model achieved the best goodness-of-fit for each modelling for the execution time random variable. Stronger shades reflect higher values in the respective cell.}
    \label{fig:win-count}
    \includegraphics[width=\linewidth]{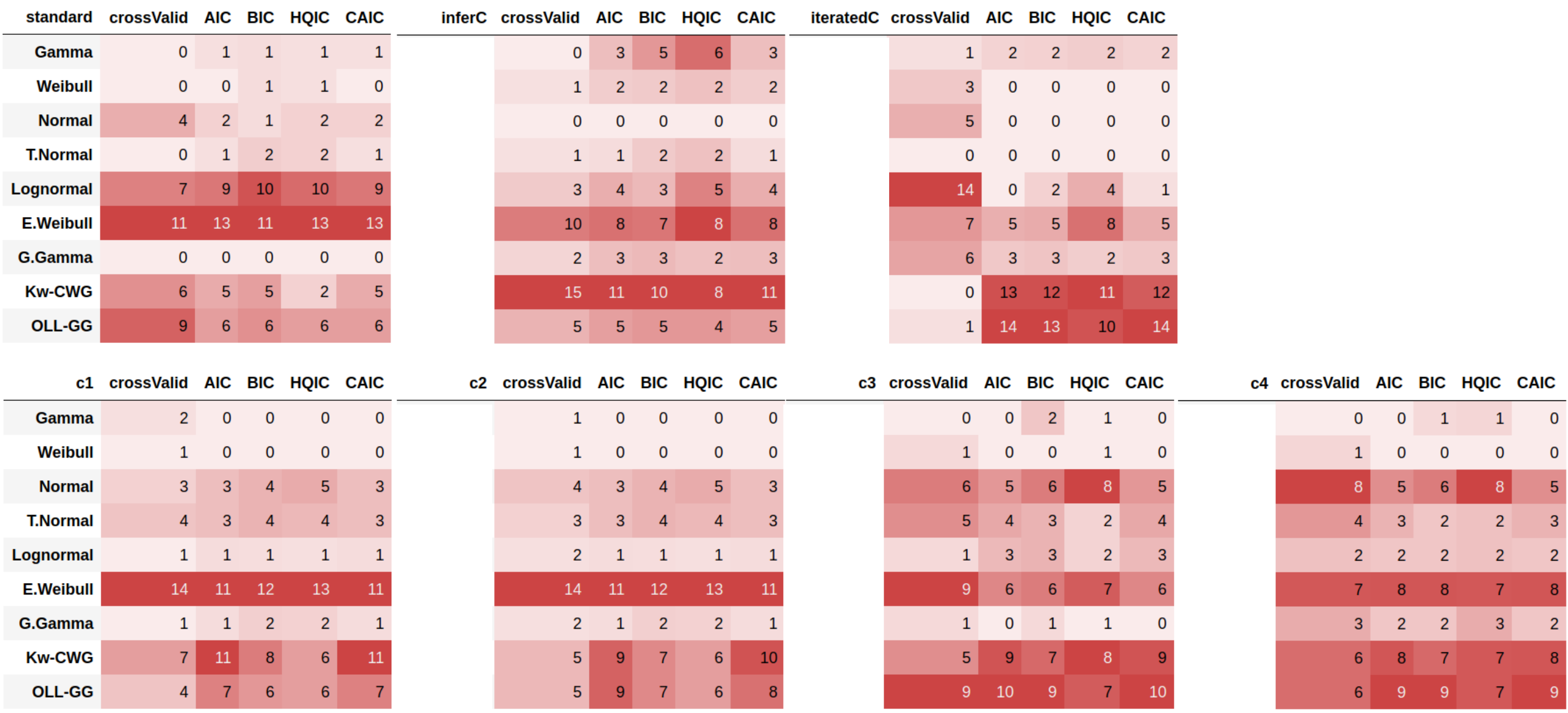}
    \caption{Number of times each distribution model achieved the second best goodness-of-fit.}
    \label{fig:win2-count}
\end{figure}

\sigla*{AIC}{Akaike information criterion} \sigla*{CAIC}{consistent Akaike information criterion} \sigla*{HQIC}{Hannan–Quinn information criterion} \sigla*{BIC}{bayesian information criterion}

To begin with, an overview of the experimental results is shown in Figures \ref{fig:win-count} and \ref{fig:win2-count}. In Chapter \ref{chapter:inf-location-parameter} seven inference methods for a random variable were presented, which here we apply to the variable $T$ of execution times. For each of these methods and each distribution family, we fitted the distribution on all experimental data and then counted how often each distribution family achieved the best fitting. Thus, the figures show the performance of each pair method-distribution, in terms of the metrics mentioned previously (recall that lower is better).

Points worth noticing in these results are:
\begin{enumerate}[nosep,noitemsep]
    \item the OLL-GG family is the best model in most metrics and inference methods;
    \item the Kw-CWG performs well on the $c_k$ and \texttt{iteratedC} inference methods (cross validation in \texttt{iteratedC} is very peculiar, so it can be disregarded here);
    \item the exponentiated Weibull achieved very good results in the standard method, meaning it probably has a more regular likelihood surface, even for data with a location parameter with large value;
    \item the truncated normal and lognormal distributions achieve the best results with a reasonable frequency;
    \item the exponentiated Weibull is the second best model in most inference methods and metrics;
    \item there is a clear transition on the second places when looking at them in the order $\hat{c}_1$ to $\hat{c}_4$, where the ``win count'' seems to spread out over all models, which might be caused by a worsening of the exponentiated Weibull on $\hat{c}_3$ and $\hat{c}_4$.
\end{enumerate}

\begin{figure}[htb]
    \centering
    \includegraphics[width=0.85\linewidth]{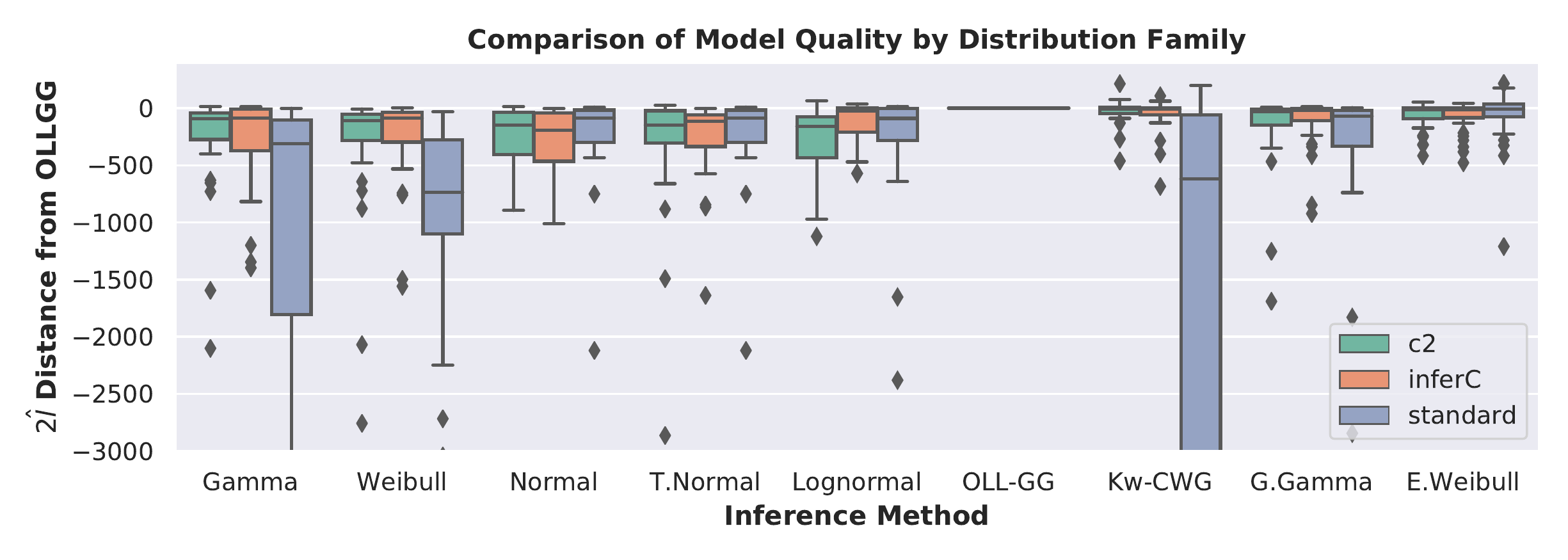}
    \includegraphics[width=0.85\linewidth]{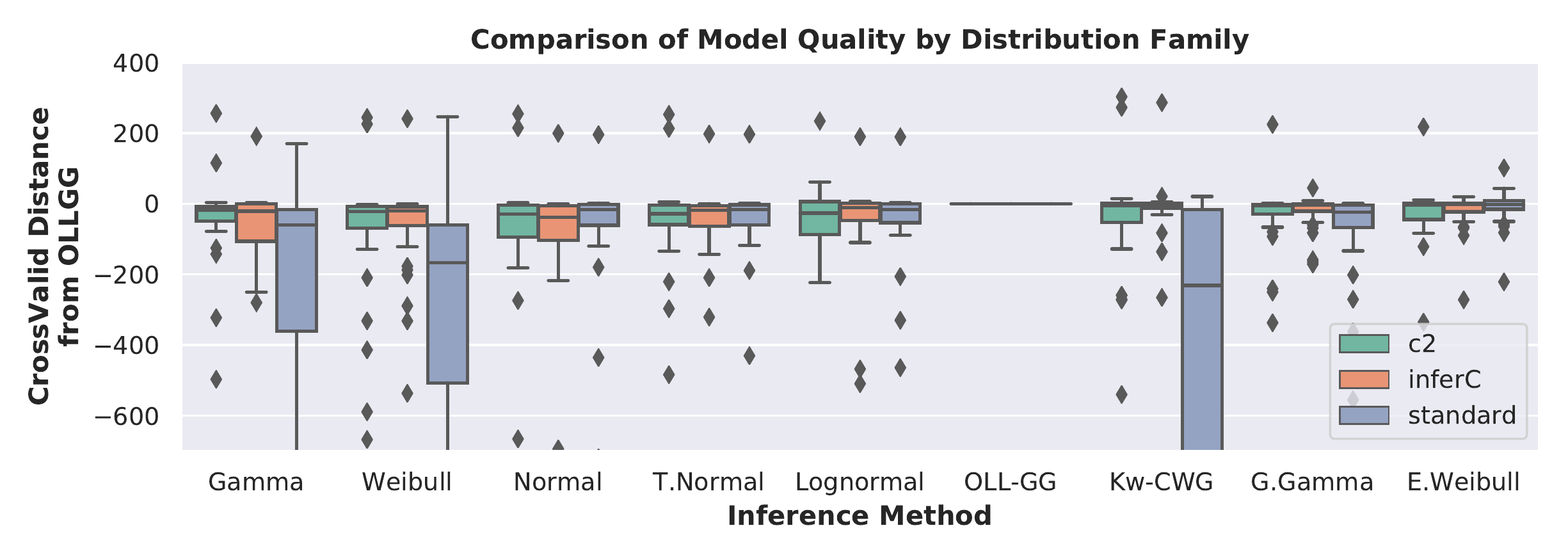}
    \caption{Distance, in achieved log-likelihood (top) and cross validated log-likelihood (bottom), from the OLL-GG model in each of the 37 sample sets.}
    \label{fig:2l-by-distribution}
\end{figure}

An alternative view of the results is shown in Figure \ref{fig:2l-by-distribution}. Here we focus on two metrics, the log-likelihood $-2 \hat{l}$ and its cross validated version. For each sample set we measure the log-likelihood for a distribution family, and take its distance from the log-likelihood obtained by the OLL-GG, which displays the best results according to the already mentioned Figures \ref{fig:win-count} and \ref{fig:win2-count}. A value of $0$ here shows an equal performance; a positive value, a better performance. This process produces $37$ values for each inference method and distribution family, which are presented as boxplots in Figure \ref{fig:2l-by-distribution}. Since some inference methods resulted in very similar boxplots, we removed these ``duplicates''. The figures show that the Kw-CWG, generalized gamma and the exponentiated Weibull share very similar log-likelihoods with the OLL-GG, except in the standard method. It is also remarkable that the performance of Kw-CWG did not degrade a lot in the cross validated metric, which was not expected since it has more parameters and thus is more prone to overfitting; it could be that although it has more parameters, many combinations of these parameters lead to the same density function, meaning the space of representable functions is not 5-dimensional. Although the generalized gamma had fairly good results, it is more dispersed than the exponentiated Weibull and its medians are slightly more negative. Overall, the exponentiated Weibull seems to be a good choice with low number of parameters; for the complex models, there is still some doubt on whether to choose Kw-CWG or OLL-GG, though the latter is favored for having less parameters.

\begin{figure}[htb]
    \centering
    \includegraphics[width=0.8\linewidth]{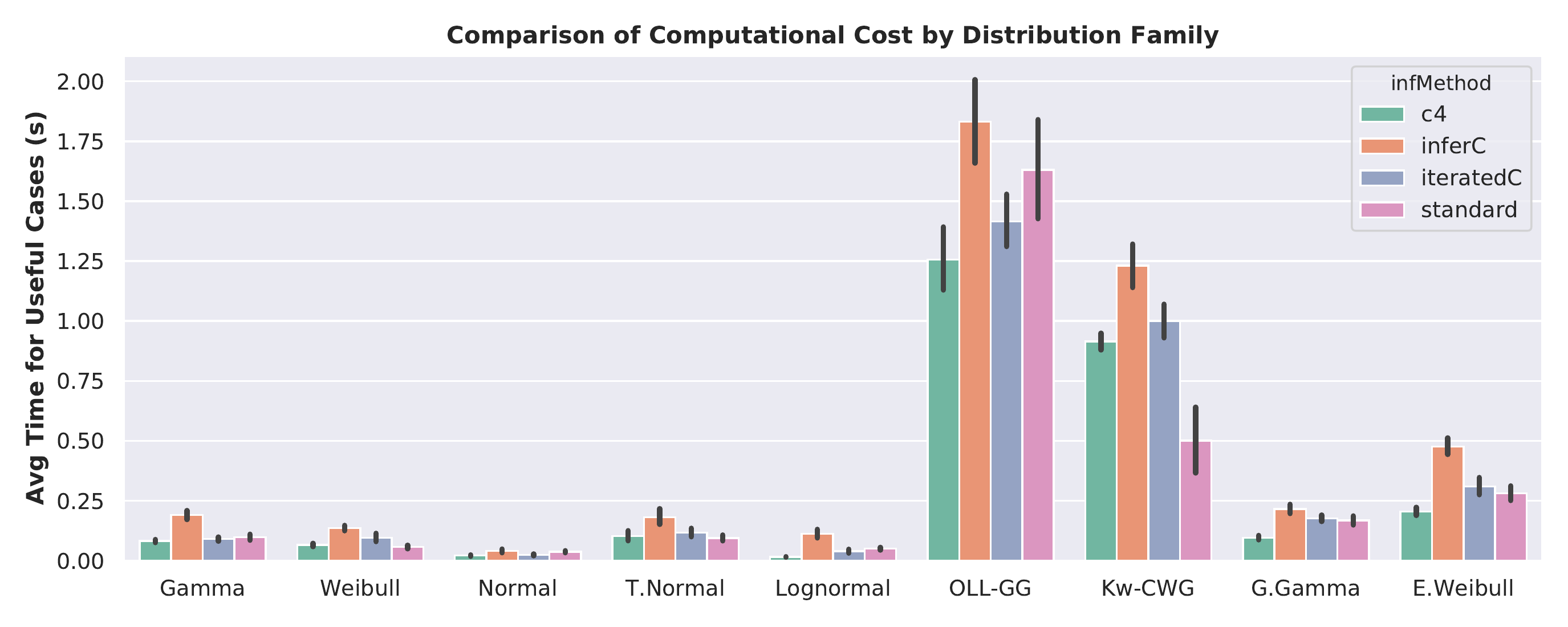}\\[0.4em]
    \includegraphics[width=0.8\linewidth]{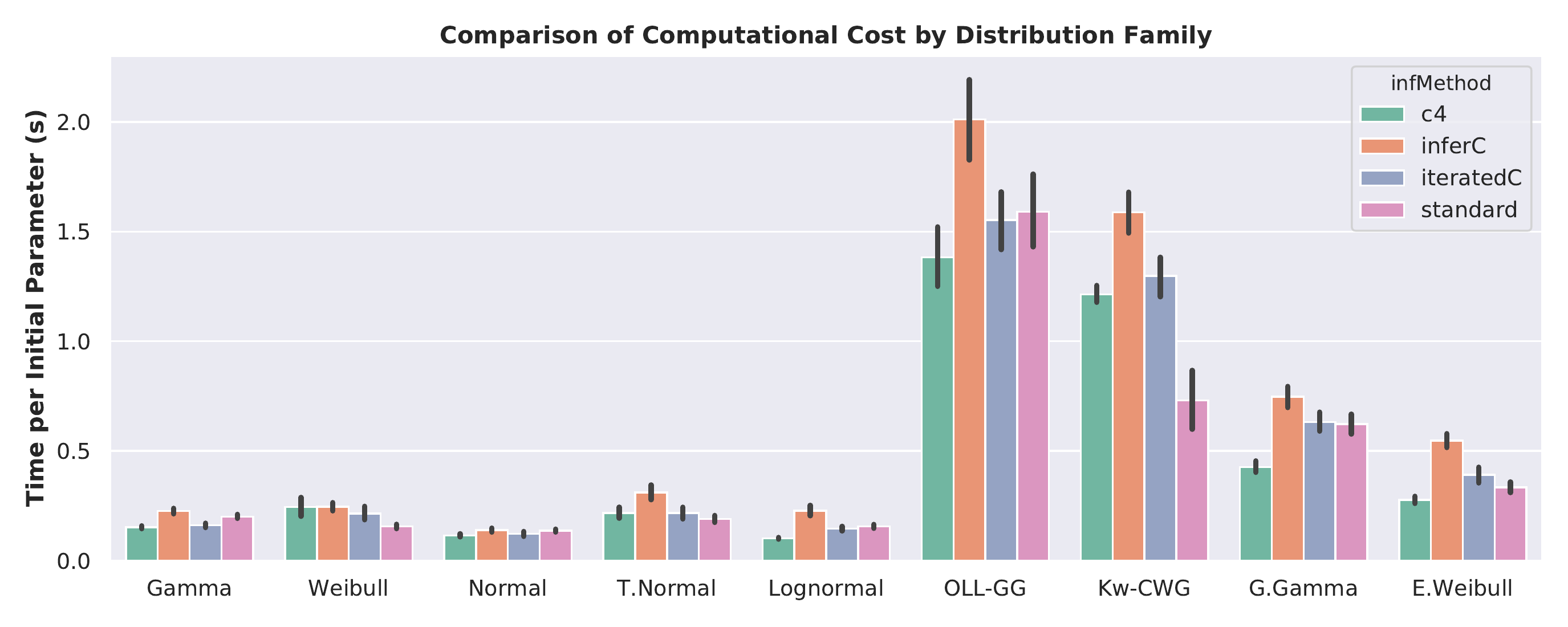}
    \caption[Comparison of execution time per MLE optimization for each distribution family, considering some inference methods.]{Comparison of average execution time over all initial parameters (bottom) and over the ones that led to convergence (top), distinguished by distribution family and considering four inference methods.  Error bars show an interval of $95\%$ confidence obtained by bootstrapping.}
    \label{fig:exec-time-distributions}
\end{figure}

So far we have only answered the question of which distribution families can model execution times more accurately. However, in the context of cloud computing, computational cost also matters a lot. Figure \ref{fig:exec-time-distributions} shows a comparison between the distribution families in this aspect\footnote{Experiments performed in the Euler supercomputer \href{http://www.cemeai.icmc.usp.br/Euler/index.html}{www.cemeai.icmc.usp.br/Euler/index.html} using an Intel Xeon E5-2680v2 2.8 GHz and 128 GB of RAM.}. Recall that MLE optimization is done as many times as there are entries in the grid of initial parameters. Similar to what was discussed in Section \ref{sec:min-estimator-experiments}, we again consider the average optimization time over all initial parameters and over the initial parameters that led the optimization to converge.

The figure shows that there is a significant difference in computational cost between OLL-GG and Kw-CWG compared with \sigla{E. Weibull}{exponentiated Weibull} and \sigla{G. gamma}{generalized gamma} distributions. This was expected due to the difference in their number of parameters, although we expected Kw-CWG (5 parameters) to be slower than OLL-GG which has 4 parameters. It could be that the extra parameters in the Kw-CWG actually make the optimization surface easier to navigate around.
One detail here is that more parameters imply that the experimenter will naturally need a larger grid of initial parameters, usually growing exponentially in size since normally one would assign initial values for each parameter $\theta_i$, and the grid would be composed of all combinations of initial values for $\theta_1, \dots, \theta_n$. Therefore, the disparity shown in Figure \ref{fig:exec-time-distributions} between E. Weibull and G. gamma with Kw-CWG and OLL-GG actually means that in practice inference will be much faster when using the former two.

Although OLL-GG was slightly slower than Kw-CWG, it should still be faster in practice due to the same reasons given above, i.e. the grid of initial conditions will be smaller for the OLL-GG since it has less parameters. On top of that, OLL-GG demonstrates a better capability to model execution times, as demonstrated in Figures \ref{fig:boxplots-inf-method-quality} and \ref{fig:win-count}. Because of these advantages, we choose OLL-GG to be our choice for modelling execution times whenever a higher number of parameters is deemed adequate. If an option with low number of parameters is desired, we recommend the exponentiated Weibull, as it led to slightly more stable likelihood values in Figure \ref{fig:2l-by-distribution} than the generalized gamma.

\rev{For the convergence of CLT see \url{https://www.google.com/search?q=central+limit+theorem+rate+of+convergence}}

\rev{Some graphics above should show information criteria instead of likelihood values.}

\rev{\section{Discussion of Results}}

\chapter{Conclusion}
\label{chapter:conclusion}

In this monograph we have investigated the problem of modelling the execution time of programs as random variables. In particular, we tried to assess the characteristics of the distributions underlying execution times, and sought a distribution family that can model reasonably well any of the possible distributions of execution times. We argued that our conclusions should hold as long as the execution path suffers little variation on each execution of the program (see Section \ref{sec:problem-program-inputs}), and the machine is sufficiently idle (see Section \ref{sec:problem-concurrent-processes}). One important observation allowed by the experiments was that execution times can have distributions with heavy tails on either side, and this can cause significant impact in scheduling decisions.

Problems arose when performing maximum likelihood estimation, which led to a whole investigation on how to perform inference when there are multiple phenomena being observed (Chapter \ref{chapter:inf-location-parameter}). Ideally, we wanted inference to work without having to set a grid of initial values for each phenomenon, and we found that this could be done by considering the underlying random variable as having a location parameter. We have proposed different inference methods, each with different intuitions and theoretical foundations, and the experiments performed so far have showed that they work quite well.

The main contributions of this work up to this point are: 1) the use of statistics tools to model the randomness of execution times was investigated, and its limitations, requirements and other considerations were extensively discussed in Section \ref{sec:suitability-of-prob-models}; 2) we proposed solutions to the sub-problem of performing inference on many different random variables in an easier way, and performed experiments showing their potential; 3) a large number of experiments were carried out in varied machines to test different probability models on modelling execution times, and the results have been exposed in Section \ref{sec:experiments-results}; 4) also in that Section, experimental data was analyzed and we concluded that, among the distribution families tested, the odd log-logistic generalized gamma (OLL-GG) and exponentiated Weibull are the most suitable to model execution times, though the latter requires a smaller grid of initial parameters and thus less computational power; and 5) we managed to publish in the official R package repository implementation of the generalized gamma, OLL-GG and Kw-CWG probability distributions, which is available at \href{https://cran.r-project.org/web/packages/index.html}{cran.r-project.org} under the names of \texttt{elfDistr}, \texttt{ggamma} and \texttt{ollggamma}. All code is available in this project's Github \url{https://github.com/matheushjs/ElfProbTET}, and more details about experimental data and their histograms can be found in the project's web page \url{https://mjsaldanha.com/sci-projects/3-prob-exec-times-1/}.


\section{Limitations and Future Work}

The main limitation of this work is the lack of machines of varied types so that experiments could also be carried out on them. Our initial hypothesis that a single distribution family could model all execution times in any machine cannot be considered as proved precisely due to the lack of different machines. Also, it would be ideal to test more distribution families, especially some that can fit bimodal distributions, as bimodality was observed in some experimental data. This project excluded programs that make use of coprocessors such as GPUs, which could also be investigated. Finally, the three chosen experimental programs were assumed to be representative of the set of all possible programs, under the argument that they exercised the three main different components of a computer: CPU, memory and disk; future work will validate such assumption empirically.

We also expect to investigate the relation of the distributions with the execution path of the programs; for now, it is hypothesized that Lyapunov's central limit theorem \citep{billingsley2008probability} can be used to justify that long execution paths tend to have distributions more similar to a normal.

Finally, the inference methods proposed in Chapter \ref{chapter:inf-location-parameter} lack more theoretical analysis. For example, we would like to prove that all $\hat{c}_k$ estimators are consistent, which does not seem difficult if we use concepts of convergence in probability. We also need to perform experiments with varying sample sizes, which we plan to do in order to publish a paper on the subject.

\section{Acknowledgements}

The author thanks prof. Ricardo Marcacini for the valuable suggestion of including the cross validation metric in our experiments. I also thank the LaSDPC and BioCom laboratories for the computational and other resources, as well as CeMEAI (FAPESP grant 2013/07375-0) for providing access to their supercomputer.


\bookmarksetup{startatroot}%

\postextual

\bibliographystyle{apalike}
\setlength{\bibhang}{0pt}
{ \linespread{1.0}
\bibliography{references}
}



\begin{apendicesenv}

    

\end{apendicesenv}


\begin{anexosenv}


\end{anexosenv}

\end{document}